\documentclass[aps,prb,twocolumn,amsmath,amssymb,superscriptaddress,floatfix]{revtex4}
\pdfoutput=1
\usepackage[colorlinks=true,linkcolor=blue,citecolor=blue,urlcolor=blue]{hyperref}

\usepackage{graphicx}
\usepackage{graphics}
\usepackage[export]{adjustbox}
\usepackage{bm}
\usepackage{amsfonts}
\usepackage{amssymb}
\usepackage{amsmath}
\usepackage{wasysym}
\usepackage{bbold}
\usepackage{stmaryrd}
\usepackage[usenames]{color}
\usepackage{colordvi}
\usepackage{units}
\usepackage{bbm}
\usepackage{booktabs}
\usepackage{array}
\usepackage{placeins}
\usepackage{epsfig}
\usepackage{lscape}
\usepackage{changes}
\usepackage{braket}
\usepackage{tabularx}
\newcolumntype{L}[1]{>{\raggedright\arraybackslash}p{#1}} 
\newcolumntype{C}[1]{>{\centering\arraybackslash}p{#1}} 
\newcolumntype{R}[1]{>{\raggedleft\arraybackslash}p{#1}} 

\newcommand{\be}{\begin{equation}}
\newcommand{\ee}{\end{equation}}
\newcommand{\beqn}{\begin{eqnarray}}
\newcommand{\eeqn}{\end{eqnarray}}

\bibstyle{apsrev.bib}

\begin{document}

\title{Correlated clusters of closed reaction centers during induction
of intact cells of photosynthetic bacteria}
\author{P\'eter Mar{\'o}ti}
\thanks{Corresponding author\\
Address: Rerrich Béla tér 1. Szeged H-6720 Hungary\\
phone: +36-62-544-120}

\email{pmaroti@physx.u-szeged.hu}
\affiliation{Department of Medical Physics and Informatics, Szeged University, H-6720 Szeged, Hungary}
\author{Istv\'an A. Kov\'acs}
\email{istvan.kovacs@northwestern.edu}
\affiliation{Department of Physics and Astronomy, Northwestern University, Evanston, IL 60208-3112, USA}
\affiliation{Wigner Research Centre for Physics, Institute for Solid State Physics and Optics, H-1525 Budapest, P.O. Box 49, Hungary} 
\affiliation{Department of Network and Data Science, Central European University, Budapest, H-1051, Hungary}
\author{Mariann Kis}
\email{kis.mariann.m@gmail.com}
\affiliation{Department of Medical Physics and Informatics, Szeged University, H-6720 Szeged, Hungary}
\author{James L. Smart}
\email{jsmart@utm.edu}
\affiliation{Department of Biological Sciences, University of Tennessee at Martin, Martin, TN 38238 USA}
\author{Ferenc Igl{\'o}i}
\email{igloi.ferenc@wigner.mta.hu}
\affiliation{Wigner Research Centre for Physics, Institute for Solid State Physics and Optics, H-1525 Budapest, P.O. Box 49, Hungary}
\affiliation{Institute of Theoretical Physics, Szeged University, H-6720 Szeged, Hungary}
\date{\today}

\begin{abstract}
Antenna systems serve to absorb light and to transmit excitation energy to the reaction center (RC) in photosynthetic organisms. As the emitted (bacterio)chlorophyll fluorescence competes with the photochemical utilization of the excitation, the measured fluorescence yield is informed by the migration of the excitation in the antenna. In this work, the fluorescence yield concomitant with the oxidized dimer (P\textsuperscript{+}) of the RC were measured during light excitation (induction) and relaxation (in the dark) for whole cells of photosynthetic bacterium \textit{Rhodobacter sphaeroides} lacking cytochrome \emph{c}\textsubscript{2} as natural electron donor to P\textsuperscript{+} (mutant \textit{cycA}). The relationship between the fluorescence yield and P\textsuperscript{+} (fraction of closed RC) showed deviations from the standard Joliot-Lavergne-Trissl model: 1) the hyperbola is not symmetric and 2) exhibits hysteresis. These phenomena originate from the difference between the delays of fluorescence relative to  P\textsuperscript{+} kinetics during induction and relaxation, and in structural terms from the non-random distribution of the closed RCs during induction. The experimental findings are supported by Monte Carlo simulations and by results from statistical physics based on random walk approximations of the excitation in the antenna. The applied mathematical treatment demonstrates the generalization of the standard theory and sets the stage for a more adequate description of the long-debated kinetics of fluorescence and of the delicate control and balance between efficient light harvest and photoprotection in photosynthetic organisms. 
\end{abstract}

\pacs{}

\maketitle

\section{Introduction}
Photosynthesis is responsible for the genesis, development and
regulation of all forms of life on the Earth by using the ultimate free
energy source of the sun. The conversion of (sun)light to chemical
energy is initiated by the absorption of the photons in the closely
packed network of protein-pigment complexes (antenna) followed by
funneling of the excitation energy (exciton) to a specially organized
(B)Chl dimer (P) in the reaction centers (RC)\cite{Mirkovic_et_al_2016}.
Here an electron is stripped from P (P$\to$P\textsuperscript{+}) converting
the energy of the exciton into chemical (redox) energy of
P/P\textsuperscript{+}. The electron is transferred via the primary
quinone acceptor Q\textsubscript{A} to the secondary quinone acceptor
Q\textsubscript{B} producing a series of transient charge separated
states (P\textsuperscript{+}Q\textsuperscript{--}). While
Q\textsubscript{A} can accept one electron only, Q\textsubscript{B}
performs two-electron chemistry: by binding two protons and forming
reduced quinone QH\textsubscript{2}, it is exchanged for an oxidized
quinone from the quinone pool in the membrane\cite{Maroti_and_Govindjee_2016,Maroti_2019a,Maroti_2019b}.

To describe the functional cooperation of the antenna pigments in light collection, the loose concept of the photosynthetic unit (PSU) was introduced\cite{Franck_and_Teller_1938}.  According to the present
knowledge, the structure of the PSU of photosynthetic bacteria can be
identified as the core complex including the photochemical RC and the
closely attached light-harvesting (core) antenna (LH1, B870 in
\emph{Rhodobacter (Rba.) sphaeroides}) together (if exists) with the
peripheral antenna (LH2, B800-850 complex in \emph{Rba. sphaeroides})
loosely arranged in the photosynthetic membrane (fluid-mosaic-membrane
model)\cite{Niederman_2016}.  The energy of light harvested by LH2 is
transferred to LH1, which directs these excitations (excitons) to an
open RC. Here the migration of the exciton in the antenna is terminated,
as it is trapped by the RC, which then becomes closed (photochemically
incompetent). Another excition visiting the closed RC can be redirected
to an open RC. The peripheral LH2 controls the exciton transfer out of
the PSU and acts as a sort of insulator between the PSUs (it can
decrease the rate of the inter-unit transfer of the excitons). The
exciton is able to visit several PSUs during its lifetime. The search
for utilization of the exciton by photochemistry (charge separation)
competes with loss by fluorescence emission. This competition is
manifested in an inverse relation between energy trapping in RCs and
fluorescence yield of the light harvesting bacteriochlorophylls, first
recognized in photosynthetic purple bacterium by Vredenberg and Duysens
in 1963\cite{Vredenberg_and_Duysens_1963}. This discovery initiated a wealth of studies on the ways and
kinetics of the RC occupation by the excitons and its correlation with
the change of the BChl fluorescence yield. As the excitions can visit
several PSUs, the first studies assumed free diffusion of excitons over
very large region (``lake model''\cite{Joliot_et_al_1973,Paillotin_1976}).
However, limitations due to the structural organization of the antenna
and to kinetic constraints restricted the number of visits to a few PSUs
(``connected units model''\cite{Lavergne_and_Trissl_1995,de_Rivoyre_et_al_2010}).

There are two distinct methods to describe the migration of the excitons
and the closure of the RCs. The first assumes homogeneous distribution
of the reactants combined with small set of reaction rate constants
\cite{Trissl_1996,de_Rivoyre_et_al_2010}. This simplified treatment has the
advantage of digestible interpretation of the experimental results by
solution of set of ordinary differential equations. The second is a more
accurate treatment of the exciton diffusion using the master equation
approach \cite{Paillotin_1976,Den_Hollander_et_al_1983,Fassioli_et_al_2009},
Monte Carlo (MC) calculations\cite{Sebban_and_Barbet_1985} or the
methods of the statistical physics. The disadvantage of this method is
the partial loss of the possibility of straightforward comparison of the
outcomes with the experimental results and of direct correspondence of
the parameters with the measurable quantities.

In the present work, the problem of excitonic connectivity in bacterial
antenna system is revisited, addressing both experimental and
theoretical aspects. The experimental section demonstrates remarkable
findings on fluorescence and absorption change kinetics both under
continuous excitation (induction) and subsequently in the dark
(relaxation). We observed 1) the enhancement of the absorption cross
section of the open RC when its neighbors were closed and 2) a
clustering of closed RCs during induction that failed during relaxation.
The theoretical section aims to provide a comprehensive toolbox to
handle the exciton migration within the organized antenna, terminating
in capture by an open RC (photochemical utilization) or waste by
fluorescence emission. This theory relies on a random walk of the
excitons on a two-dimensional square lattice and permits temporal
evaluation of the state of the RC (open vs. closed) and of the
fluorescence both upon continuous excitation and in the dark. The method
sets the stage to understand the observed complexity of the fluorescence
induction and relaxation kinetics in purple bacteria \cite{Asztalos_et_al_2015,Maroti_2016} 
and in PSII of higher plants\cite{Kupper_et_al_2019}. A
direct comparison is made with the results from the homogeneous kinetic
(Joliot) model. In the future, the theoretical treatment can be extended
for more accurate description of the exciton diffusion in various types
of and more realistic models of antenna and RC organization supported by
recent electron and atomic force microscopy\cite{Niederman_2013}.

\section{Results}
\label{sec:results}

\subsection{Experimental Results}
\label{sec:experimental_results}

\emph{Kinetics of fluorescence and absorption change}

\begin{figure}
\includegraphics[width=1.\columnwidth, center]{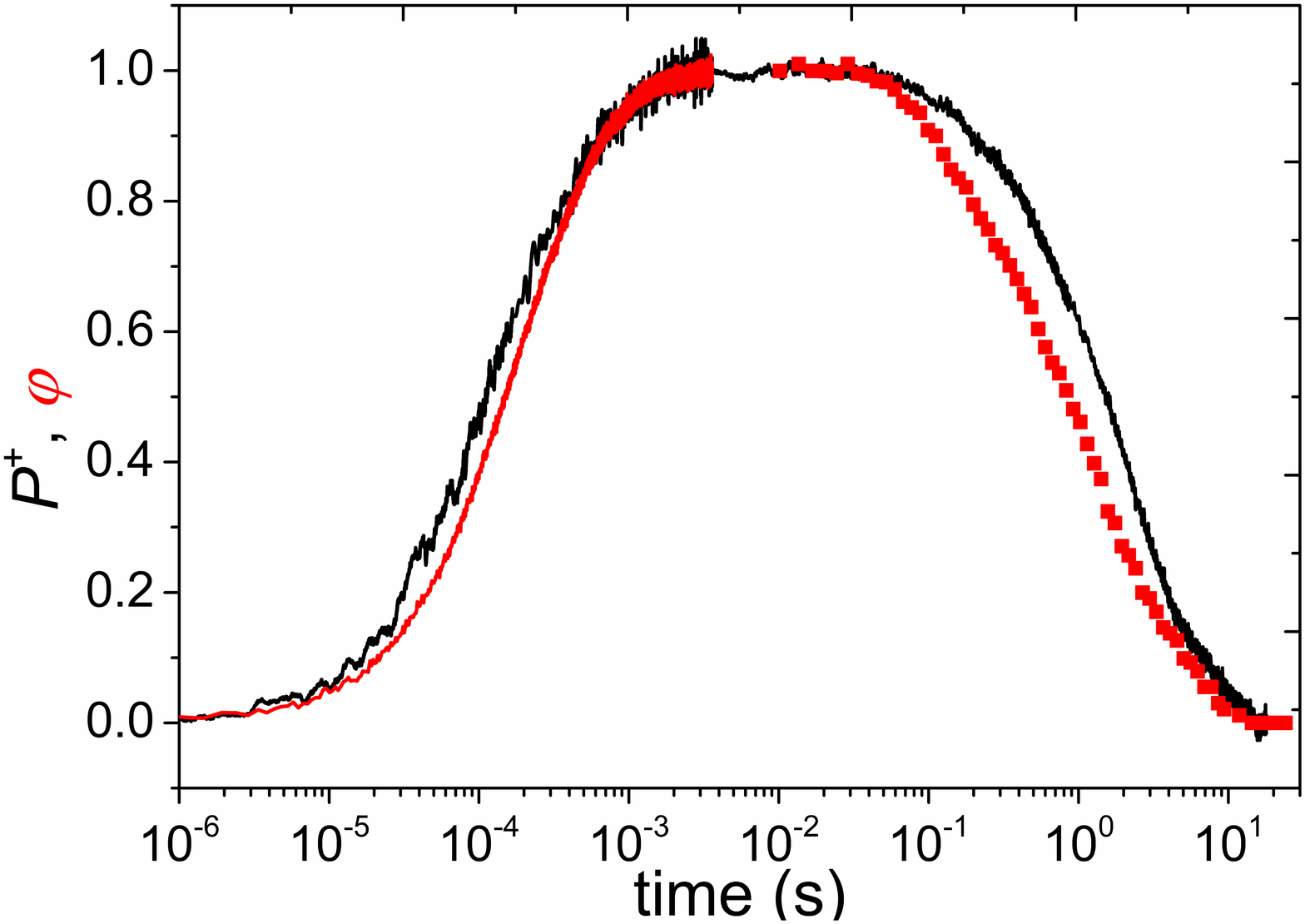}
\caption{Kinetics of fluorescence yield ($\varphi$) and
absorption changes of oxidized dimer (P\textsuperscript{+}) during induction
and relaxation of whole cells of cytochrome \emph{c}\textsubscript{2}
less mutant of purple photosynthetic bacterium \emph{Rba. sphaeroides}.
Both $\varphi$ and P\textsuperscript{+} are normalized to their maximum
values.
\label{fig_01}}	
\end{figure}

One of the most convenient methods to study the excitonic coupling among
the PSUs is the measurement of the kinetics of the induction and
subsequent relaxation of the yield of fluorescence emitted by the BChl
antenna. The rise in fluorescence was detected upon laser diode
excitation and the decay was monitored by a series of short laser diode
probing flashes. A typical experiment is shown in Fig. \ref{fig_01} on whole cells
of the cyt \emph{c}\textsubscript{2} less mutant of photosynthetic
purple bacterium \emph{Rba. sphaeroides} (\textit{cycA} strain). As the variant
lacks natural electron donors to P\textsuperscript{+}, single turnover
of the RC is assured upon excitation whose duration is less than the
P\textsuperscript{+}Q\textsubscript{A}\textsuperscript{--}$\to$PQ\textsubscript{A}
charge recombination time ($\sim$100 ms). The fluorescence
(induction) followed the PQ\textsubscript{A} $\Rightarrow$
P\textsuperscript{+}Q\textsubscript{A}\textsuperscript{--}
photochemistry with rise time inversely proportional to the exciting
light intensity. A rise time in the submillisecond time range was selected
to avoid complications with the appearance of short lived triplet
quenchers and with the charge recombination on the microsecond and 100
ms time scales, respectively. The applied light intensity was able to
saturate the fluorescence within 5 ms. The saturated high fluorescence
state was a long lived state indicated by the slow relaxation of the
fluorescence in the dark ($\sim$ 1 s) in accordance with the
re-reduction of P\textsuperscript{+} by the
P\textsuperscript{+}Q\textsubscript{B}\textsuperscript{--}$\to$PQ\textsubscript{B}
charge recombination.

To monitor the time course of P\textsuperscript{+}, the kinetics of
absorption change at 790 nm were measured. This signal is attributable
to an electrochromic shift of the absorption band of the BChl monomers
in the RC that is induced by the dimer. The conditions were the same in
fluorescence and absorption experiments both during induction and in
relaxation. While the rise of the fluorescence was slower with respect
to absorption change in the induction, the decay of fluorescence was
faster than that of the absorption in the relaxation. As both signals
were normalized to their full amplitude, we conclude that the
fluorescence was always below the absorption. The observed difference in
the kinetics reflects the excitonic connectivity of the PSU
characterized by the Joliot parameter \emph{p}. Additionally, the
difference between the kinetic traces of fluorescence and absorption
change is larger during relaxation than during induction indicating the need for
formal use of different \emph{p} values in the light (induction) and in
the dark (relaxation).

\emph{Deviations from the Joliot model}

Due to energetic coupling of the PSUs, the particular kinetics of rise
and decay of fluorescence and oxidation of P and re-reduction of
P\textsuperscript{+} do not follow single exponential function but are
multiphasic. We will investigate the accuracy of the Joliot treatment to
describe the fluorescence kinetics.

\begin{figure}
\includegraphics[width=1.\columnwidth, center]{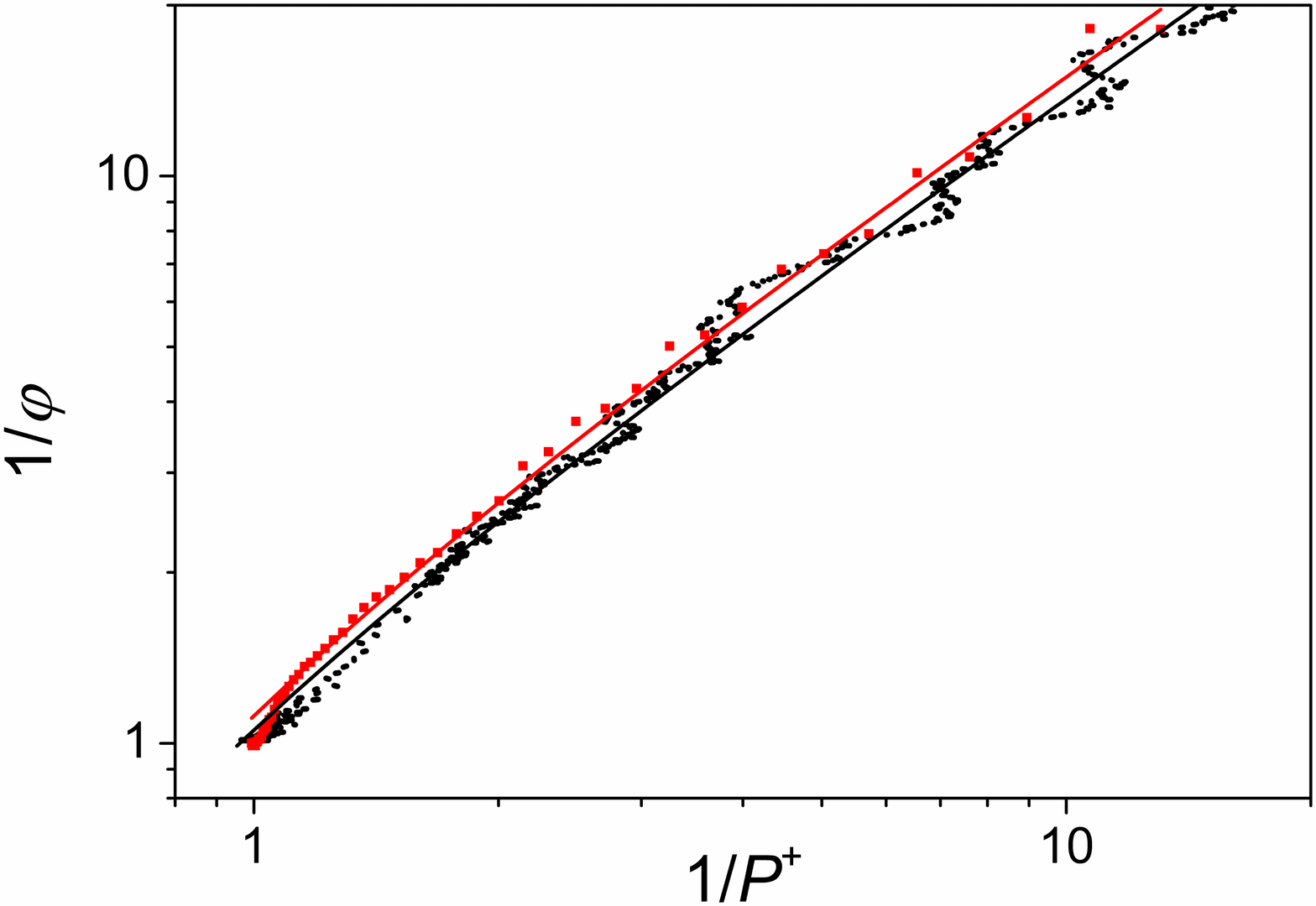}
\caption{Double reciprocal plots of the fluorescence yield
($\varphi$) and complementary area (\emph{C}) above the fluorescence rise
(induction) or below the fluorescence drop (relaxation), respectively.
Systematic deviations can be observed from straight lines predicted by
the Joliot model. Note, that the straight lines are curved in
logarithmic scales.
\label{fig_02}}	
\end{figure}

According to the Joliot theory, the yield of the fluorescence $\varphi$ can be
expressed by the complementary area of the
fluorescence \emph{C}(\emph{t})
\be
\frac{1}{\varphi(t)}=\frac{1}{k_I (1-p)} \frac{1}{C(t)}-\frac{p}{1-p}\;,
\label{J1}
\ee
where \emph{k}\textsubscript{I} is the photochemical rate constant
(time-scaling factor). Eq. (1) predicts that the fluorescence data in
double-reciprocal representation of $\varphi(t)$ vs.
\emph{C}(\emph{t}) should be linear. (The complementary area is defined
as the area that is above the fluorescence rise during induction and below
the fluorescence decay during relaxation.) Although the plots in Fig.\ref{fig_02} are
close to straight lines, the measured data do not scatter randomly
around the straight line: the deviations are systematic. They are
smaller during induction and larger during relaxation of the fluorescence.

\begin{figure}
\includegraphics[width=1.\columnwidth, center]{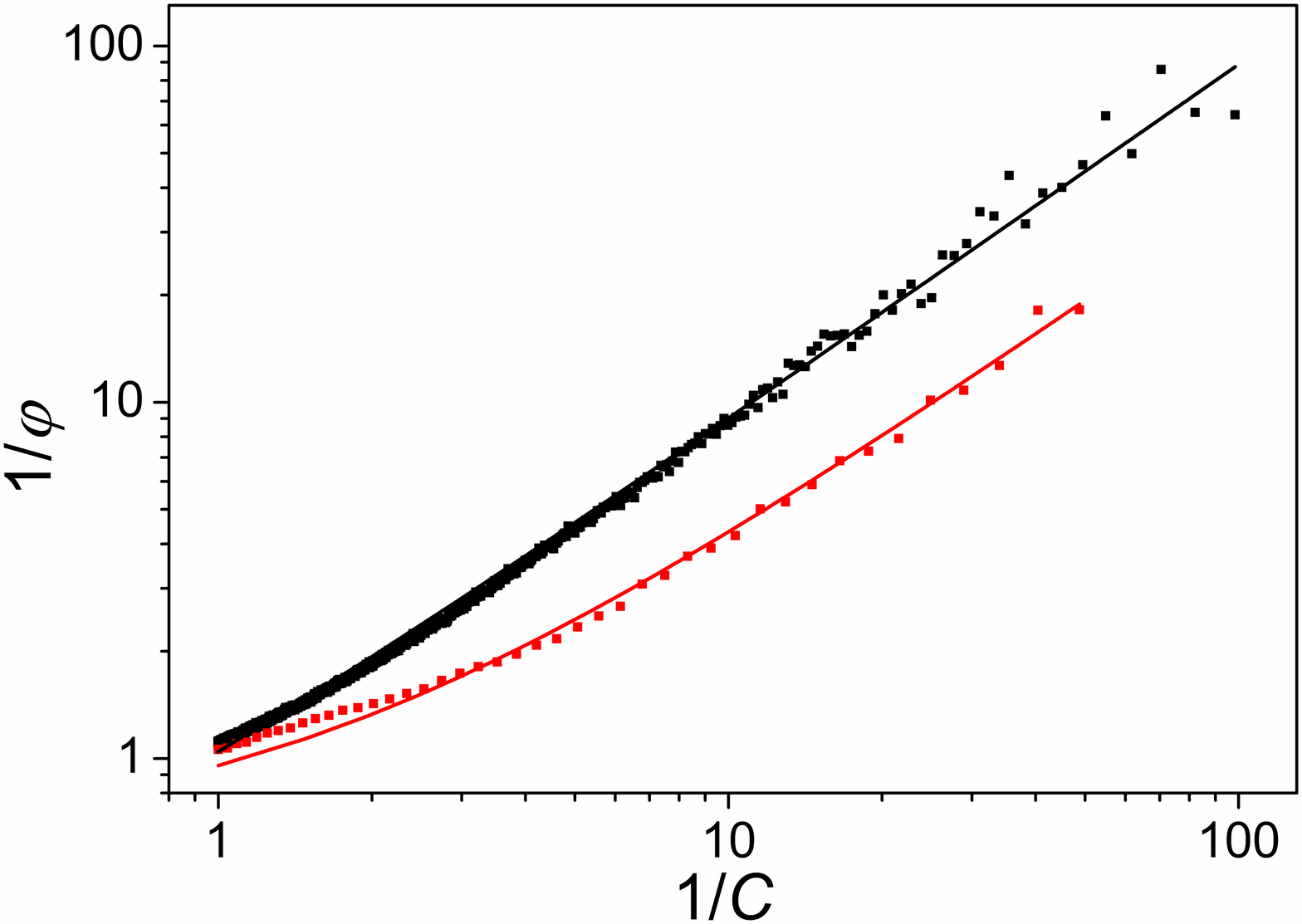}
\caption{Double reciprocal representations of the fluorescence
yield ($\varphi$) and oxidized dimer (P\textsuperscript{+}) during induction
and relaxation. The plots demonstrate the deviation from the Joliot
model that predicts straight lines (curved in the logarithmic scale).
\label{fig_03}}	
\end{figure}

Similarly, the yield of fluorescence is a hyperbolic (and not a linear)
function of P\textsuperscript{+} concentration due to the connectivity of the PSUs
characterized by the Joliot parameter \emph{p}:
\be
\frac{1}{\varphi(t)}=\frac{1}{(1-p)} \frac{1}{P^+}-\frac{p}{1-p}\;,
\label{J2}
\ee
The validity of this expression can be checked by simultaneous
measurement of fluorescence and P\textsuperscript{+} from absorption
change. In double reciprocal representation, systematic deviation from
the straight line can be observed both during induction and during relaxation
(Fig. \ref{fig_03}). The divergence is large in the vicinity of the borders (0 and
1).

\emph{Hysteresis of the $\varphi$ vs. P\textsuperscript{+} relationship}

\begin{figure}
\includegraphics[width=1.\columnwidth, center]{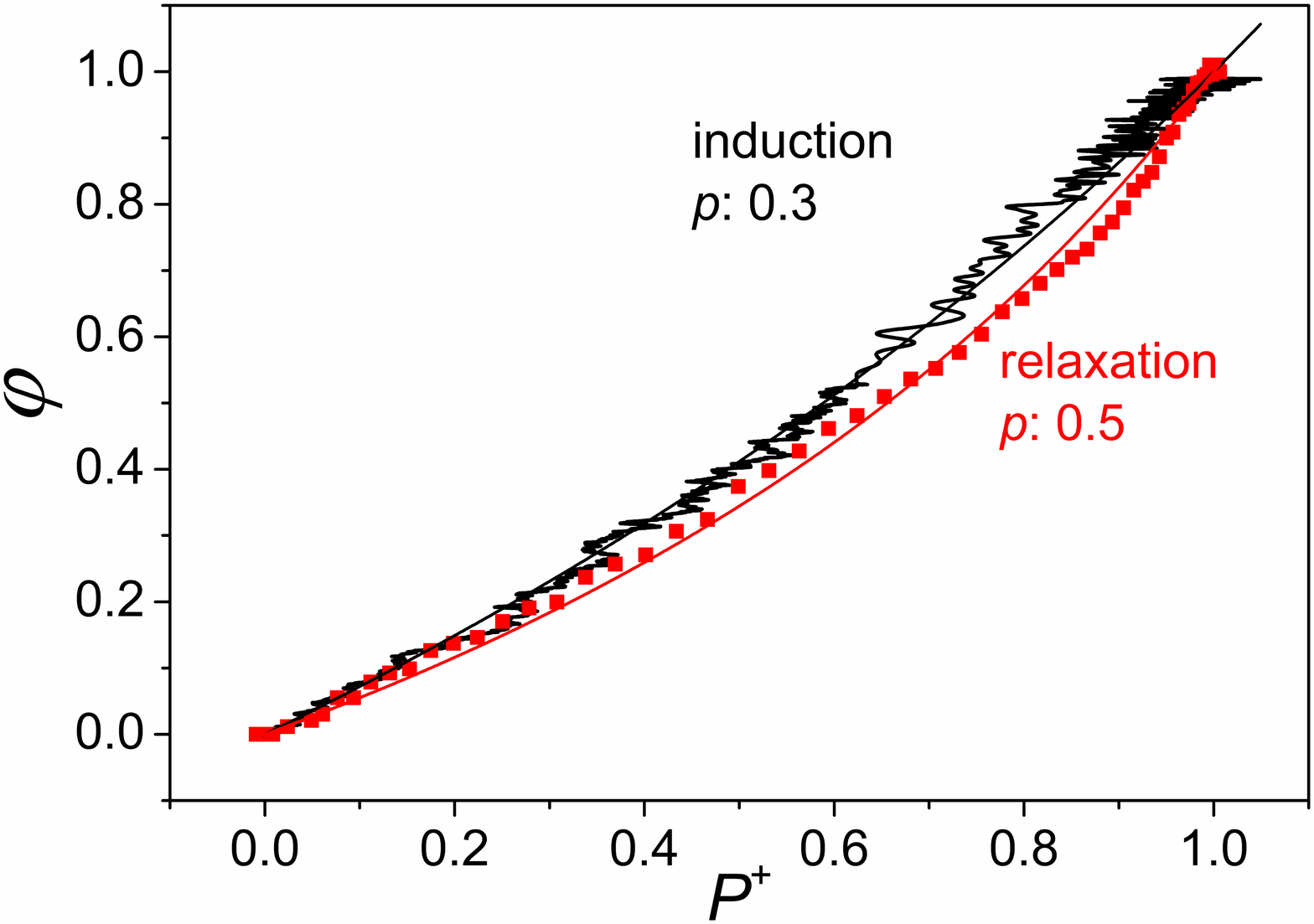}
\caption{Fluorescence yield ($\varphi$) as a function of
closure of the reaction centers (P\textsuperscript{+}) during induction
and relaxation phases obtained by comparison of the kinetics of
fluorescence with those of the oxidised dimer by elimination of the
time. The cells were harvested in the late stationary phase of their
growth (3 days after inoculation). The measured points were formally
approximated by curves derived from the Joliot model with different
\emph{p} values indicated. The straight line corresponds to $p=0$,
i.e. no connection between the PSUs. The hysteresis (the difference
between induction and relaxation) is relatively modest.
\label{fig_04}}	
\end{figure}

As we saw above, the rise in fluorescence is slower than
P\textsuperscript{+} upon illumination (induction), and the decay is
faster than P\textsuperscript{+} in the dark (relaxation). Furthermore,
these two deviations are different: the delay in fluorescence rise
relative to P\textsuperscript{+} during induction is smaller than the
increase in decay that we observed during relaxation in the dark. This
difference results in a hysteresis observable in the plot of $\varphi$
vs. P\textsuperscript{+} that can be obtained from the kinetic data
after elimination of the time variable (Fig. \ref{fig_04}). 

\begin{figure}
\includegraphics[width=1.\columnwidth, center]{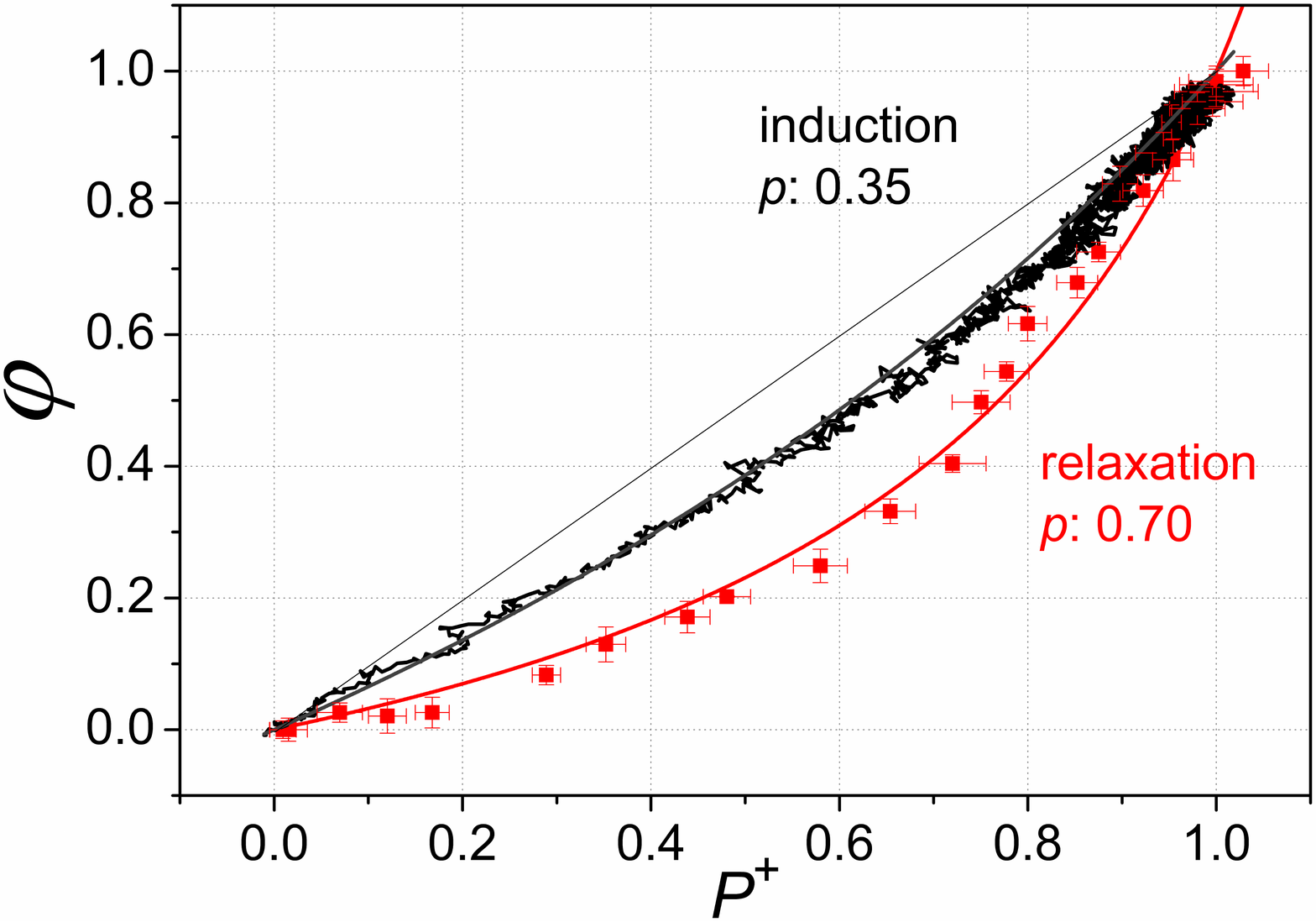}
\caption{Demonstration of large hysteresis due to the
increased difference between the kinetics of fluorescence yield and
closure of the PSU during induction and relaxation. The bacteria were
harvested in the early phase of their growth (24 hours after
inoculation). Otherwise the experimental conditions and evaluation of
the data were the same as in Fig. \ref{fig_04}.
\label{fig_05}}	
\end{figure}

The curvature of the data measured in the light is smaller than that
measured in the dark which would correspond formally to different
\emph{p} values: 0.3 during induction and 0.5 during relaxation. The magnitude of the hysteresis probably depends on the physiological condition of the bacteria, as younger (24 hours) cells demonstrate a more pronounced effect: \emph{p} = 0.35 for induction and \emph{p} = 0.70 for relaxation (Fig. \ref{fig_05}). The fit to the Joliot model is not perfect as it shows
systematic deviation from the measured points. It reflects that the
hyperbola is not symmetric i.e. it does not cut the diagonal of slope 1
at equal angles (see Eqs.
(\ref{J1}) and (\ref{J2})).

\subsection{Mathematical Results}
\label{sec:mathematical_results}

\textit{Models with exciton wandering}

A lattice-gas model is used: each RC is identified as the site within a lattice, $i=1,2,\dots,N$, which can be in two different states, ${\sigma}_i=\fullmoon$ for an open RC with a value $\sigma_i=0$ and ${\sigma}_i=\newmoon$ for a closed RC with $\sigma_i=1$. The fraction of open and closed RCs are denoted by $P_{\circ}$ and $P_{\bullet}$, respectively, such that $P_{\circ}+P_{\bullet}=1$. In the following we use the notation $P_{\bullet} \equiv x$, which is called as the order-parameter and is defined as:
\be
x=\lim_{N \to \infty}\frac{1}{N}\sum_{i=1}^N \sigma_i=\langle \sigma \rangle\;.
\ee
$x$ is actually the fraction of closed RCs, and is time dependent: $x = x(t)$. At the starting point of the induction all RCs are open, thus $x(0) = 0$ followed by continuous closure of the RCs. For long enough time, all the RCs become closed, thus $x = 1$. During relaxation, we start from a fully-closed state, and after a sufficient period of time, a fraction of the RCs $(1-x)$ will spontaneously reopen. Using an appropriate weak probing light beam, the dynamics of the system can be studied in this case, too.

In the direct process, when an incoming exciton (denoted by $\photon$) hits an open RC, say at $i_1$, the RC will become closed, which is represented graphically in the first line of Eq.(\ref{exciton}). If, however, the exciton hits a closed RC, two processes can take place. With probability $p$, the exciton visits a neighbouring RC, or with probability $(1-p)$ the energy of the exciton is dissipated by emission of a fluorescence quantum, $\textbf{pF}$, see the second line of Eq.(\ref{exciton}). 
\beqn
&\fullmoon_{i_1}& \xrightarrow{\photon} \newmoon_{i_1}\nonumber \\
&\newmoon_{i_1}& \xrightarrow[p]{\photon} \mbox{\boldmath$\photon$},\quad \newmoon_{i_1} \xrightarrow[1-p]{\photon} \textbf{pF}\;.
\label{exciton}
\eeqn
Depending on the lifetime of the exciton it can jump to a new site, say to $i_2$, and the same state-dependent processes can take place, as represented in Eq.(\ref{exciton}). This is repeated for more jumps, say on the route of closed sites $i_1 \curvearrowright i_2 \curvearrowright i_3 \curvearrowright \dots \curvearrowright i_k$, and the processes at $i_k$ are the same as in Eq.(\ref{exciton}). We note that a wandering exciton can visit the same closed RC several times.
Here we assume that the exciton can take at most $(n-1)$ jumps, so that after reaching a closed site at $i_n$ its energy is emitted as fluorescence: $\newmoon_{i_n} \xrightarrow{\photon} \textbf{pF}$.

Thus, the experimentally measured fluorescence yield is given as the sum of the contributions of the different processes:
\beqn
\varphi&=&(1-p)\langle \sigma \rangle+(1-p)p\langle \sigma_{i_1}\sigma_{i_2}\rangle + (1-p)p^2\langle \sigma_{i_1}\sigma_{i_2}\sigma_{i_3}\rangle\nonumber\\
&+&\dots+(1-p)p^{n-2}\langle \sigma_{i_1}\sigma_{i_2} \dots \sigma_{i_{n-1}}\rangle\nonumber\\
&+&p^{n-1}\langle \sigma_{i_1}\sigma_{i_2} \dots \sigma_{i_n}\rangle\;,
\label{F}
\eeqn
where the $k$-th term in the r.h.s. ($2 \le k<n$) corresponds to the process:
\be
\newmoon_{i_1} \xrightarrow[p]{\photon} \newmoon_{i_2} \xrightarrow[p]{\photon}  \dots
\xrightarrow[p]{\photon} \newmoon_{i_k} \xrightarrow[1-p]{\photon} \textbf{pF}\;,
\ee
and the $k$-site correlations are defined by the average over all possible $k$-step walks:
\be
\langle \sigma_{i_1}\sigma_{i_2} \dots \sigma_{i_k}\rangle=\frac{1}{Nz^{k-1}} \sum_{i_1 \curvearrowright i_2 \curvearrowright \dots \curvearrowright i_k}\sigma_{i_1}\sigma_{i_2} \dots \sigma_{i_k}\;.
\label{m-corr}
\ee
Here the sum over the starting position is $i_1=1,2,\dots,N$, whereas the following exciton jumps can reach $i_2,i_3,\dots,i_k=1,2,\dots z$ nearest-neighbour sites. Note, that in the sum in Eq.(\ref{m-corr}) the same site can be visited several times. In this case using the identity:
\be
\sigma_i^s=\sigma_i,\quad s=2,3,\dots\;,
\label{sigma_rel}
\ee
one obtains a set of $j$-site reduced correlations, $x_j$ with $2\le j < k$, such that
\be
G_k \equiv \langle \sigma_{i_1}\sigma_{i_2} \dots \sigma_{i_k}\rangle=\sum_{j=2}^k c_j^{(k)}x_j\;,
\label{m-corr-rel}
\ee
where $c_j^{(k)}$ is the fraction of $k$-step random walks which have visited $j$ different sites.

The general theoretical model, described above, contains i) the exciton hopping probability, $p$, ii) the possible maximal number of steps, $n$ and iii) the number of nearest-neighbours, $z$. The parameter, $p$ in the Joliot model depends on both the connectivity of the PSUs and on the RC parameters. Here, $p$ denotes the hopping probability of the exciton from closed RC to one of its neighbors. Note that the first two of these parameters ($n$ and $p$) are linked since the average number of steps of the exciton (corresponding to its life-time) is given by a combination (see Eq.(\ref{n_av}) in \ref{sec:mathematical_methods}), which has its maximal value: $\langle n \rangle =\dfrac{1-p^n}{1-p}$.

The multi-site correlations are expected to be different during induction and relaxation. During relaxation the spontaneous opening of the RCs is uncorrelated in time, so that multi-site correlations depend solely on the one-point function given by the density, $x$. For example the two-point correlation function is given by: $\langle \sigma_{i_1}\sigma_{i_2}\rangle_{\rm rel}=\langle \sigma_{i_1}\rangle\langle\sigma_{i_2}\rangle=x^2$. On the contrary, during induction the exciton makes jumps to nearest-neighbour sites, which creates short-range correlations. For example the two-point function during induction generally satisfies: $\langle \sigma_{i_1}\sigma_{i_2}\rangle_{\rm ind} \ge \langle \sigma_{i_1}\rangle\langle\sigma_{i_2}\rangle=x^2$. As a consequence $\varphi_{\rm ind} \ge \varphi_{\rm rel}$, in agreement with the experimentally observed hysteresis in the fluorescence yield.

The dynamics of closing the open RCs follows from the fact that all incoming photons that are not emitted through fluorescence will reduce the number of open RCs, thus the time-dependence of $P_{\circ}$ follows the rule:
\be
-\frac{\textrm{d} P_{\circ}}{\textrm{d} t}=\frac{\textrm{d} x}{\textrm{d} t}=1-\varphi\;,
\label{dx/dt}
\ee
where the photochemical rate constant (time-scaling factor) is set $k_I=1$.
To integrate this equation one needs to approximate the multi-site correlations in $\varphi$. In the following we recapitulate the essence of the homogeneous kinetic model (standard theory of Joliot\cite{Joliot_et_al_1973}, developed later by Lavargne and Trissl\cite{Lavergne_and_Trissl_1995}); afterwards we present our refined method, which takes into account the local topology of the RC-lattice as well as the bunching effect of the closed RCs in the induction process.

\textit{Joliot theory - mean-field approach}

\begin{figure}[ht]
	\includegraphics[width=1.0\columnwidth,angle=0]{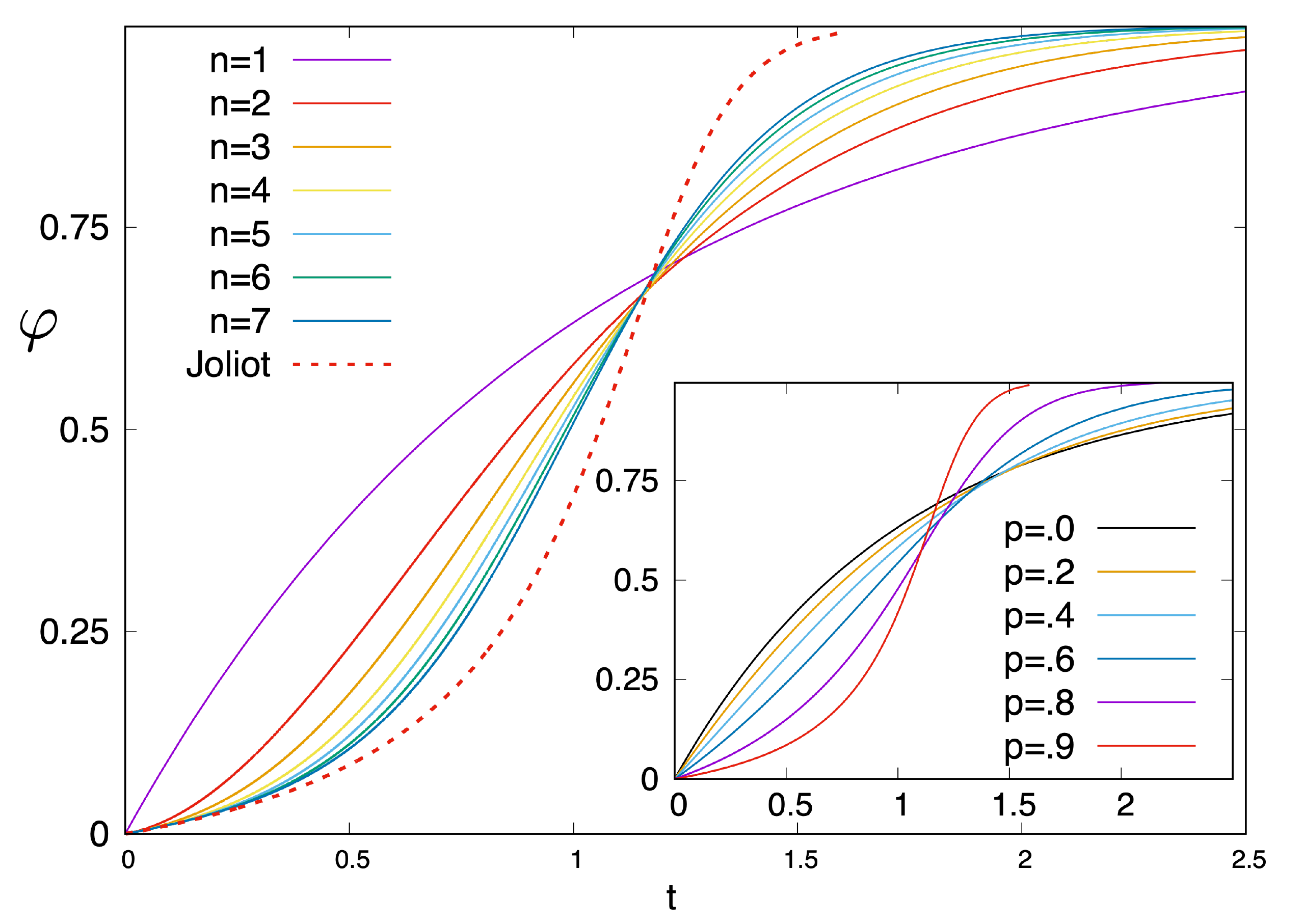}
\caption{Time-dependence of the fluorescence yield calculated by the LMF approximation for various values of $n$ at a hopping probability $p=0.9$. The result of the Joliot theory with the same $p$ is shown for comparison. Inset: Time-dependence of the fluorescence yield in the Joliot model for different values of the hopping parameter.\label{fig_2}}
\end{figure}

The Joliot theory, proposed almost half a century ago, has been used since then with great satisfaction, mainly due to its simple form\cite{Joliot_et_al_1973}. In this approach the exciton makes unlimited number of steps ($n \to \infty$) and it can hop to any lattice site, thus the local topology of the lattice is irrelevant. The multi-site correlations are factorised in terms of one-site terms: $G_k=x^k$. Due to these approximations the fluorescence yield in Eq.(\ref{F}) assumes the form of a geometric series having the sum in the simple form:
\be
\varphi=\frac{(1-p)x}{1-px}\;.
\label{Joliot_F}
\ee
which corresponds to Eq.(\ref{J1}).
Putting this into Eq.(\ref{dx/dt}) the time-evolution of $x$ is given by:
\be
\frac{{\rm d} x}{{\rm d} t}=\frac{1-x}{1-px}\;,
\label{Joliot_diff_P_p}
\ee
having the solution:
\be
px -(1-p)\ln(1-x)=t\;.
\label{Joliot_sol_P_p}
\ee
Time-dependence of the fluorescence yield is shown in the inset of Fig.\ref{fig_2} for different values of the hopping parameter, $p$. At
$p=0$ (without exciton wandering) the solution is a pure exponential $\varphi_0=x_0=1-\exp(-t)$, which is crossed by the curves with $p>0$.

\textit{Generalised mean-field models}

\begin{figure}
	\includegraphics[width=1.0\columnwidth]{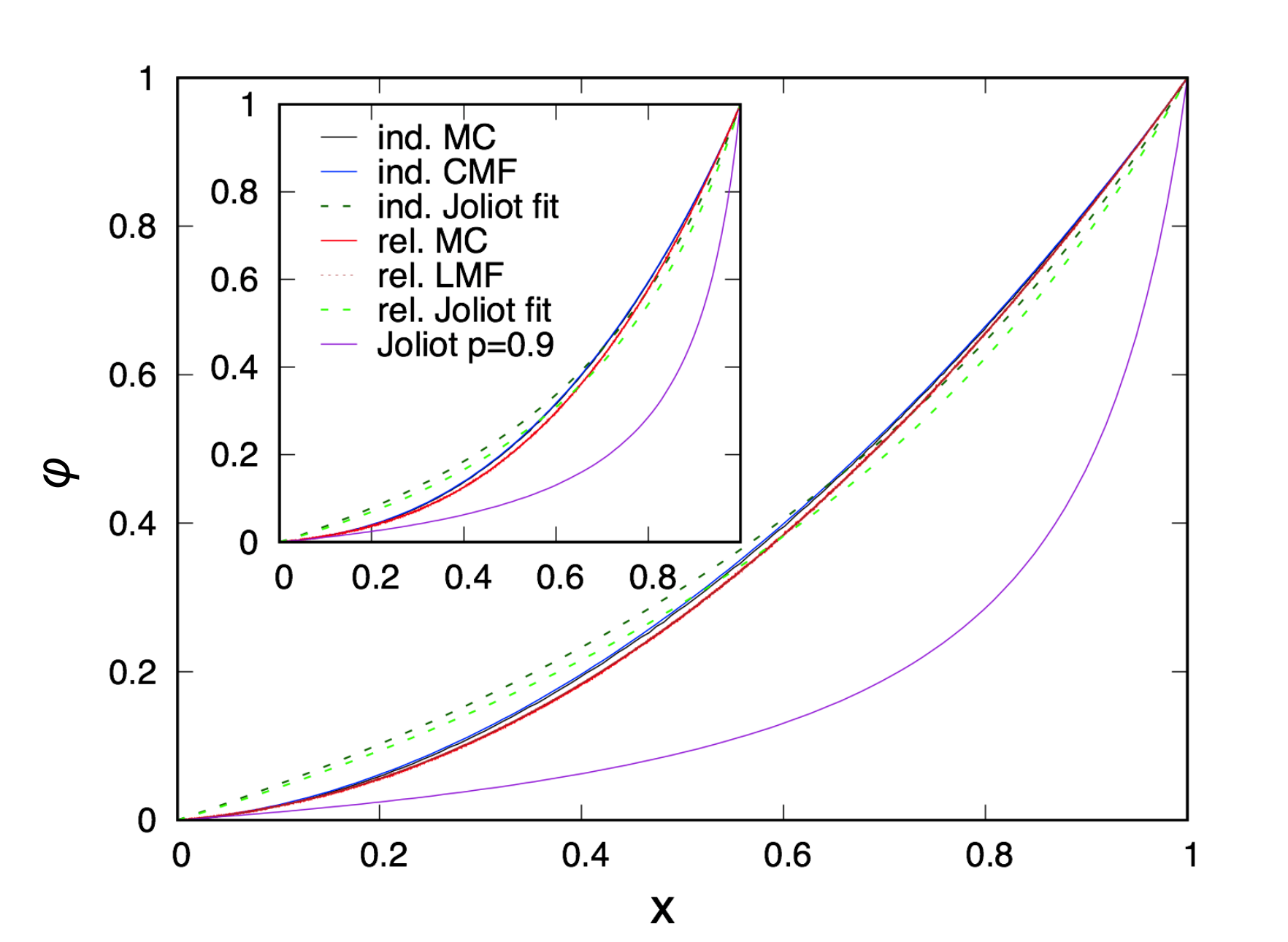}
    \caption{Fluorescence yield as a function of the fraction of closed RCs calculated during relaxation (LMF calculation and MC simulations) and during induction (CMF calculation and MC simulations) at a hopping probability $p=0.9$ for $n=2$ (main panel) and $n=3$ (inset). In both cases the best fit (with $p$) of the Joliot theory, as well as the result with a fixed $p=0.9$ is also presented.\label{fig_1}}
\end{figure}
In our approach the exciton makes a limited number of steps ($n=2,3,\dots,6,7$) and it can hop to nearest-neighbour lattice sites, thus $z$ is given by the coordination number of the lattice, see in Eq.(\ref{m-corr}). In concrete calculations we use a square lattice with $z=4$. The reduced multi-site correlations are approximated in two different ways. In the lattice mean-field (LMF) method the reduced multi-site functions are expressed as the product of one-site functions, $x_j=x^j$ ,which is expected to describe correctly the dynamics at relaxation. In the cluster mean-field (CMF) method the reduced multi-site correlations are expressed in terms of one-site, $\langle \sigma_{i_1} \rangle=x$, and two-site correlations, $\langle \sigma_{i_1}\sigma_{i_2}\rangle=x_2$, where $i_1$ and $i_2$ are nearest neighbours. In the latter method the bunching effect of closed RCs during induction is taken into account.

\textit{Lattice mean-field approach}

In the LMF approximation the reduced multi-site correlations are expressed as the products of one-site functions, see in Eq.(\ref{G_k_LMF})and the fluorescence yield is given as an $n$-th degree polynomial of $x$, which is illustrated in the insets of Fig. \ref{fig_1} for $n=2$ and $3$ at a hopping probability $p=0.9$. At $n=1$ the relation is linear, $\varphi=x$, which is modified for $n=2,3,\dots$, when exciton wandering is taken into account.

The time-dependence of the fluorescence yield is obtained by performing the integral in Eq.(\ref{int}), either analytically for $n=1,2,3$, see in Eqs.(\ref{n=1},\ref{n=2},\ref{n=3}), or numerically for $n>3$. The kinetics, $\varphi(t)$ for different values of $n$ at the hopping probability $p=0.9$ is shown in Fig. \ref{fig_2}. It is seen in this figure, that $\varphi(t)$ for $n \ge 2$ in the starting time-period is convex, which turns to concave at an inflection point. The curves with different values of $n$ cross each others approximately at the same (inflection) point. The sigmoidicity of the fluorescence induction kinetics is overestimated by the Joliot theory with the same $p=0.9$.

\textit{Cluster mean-field approach}

\begin{figure*}[ht]
\vskip -3cm
	\hskip -0.4cm
	\includegraphics[width=0.93\columnwidth]{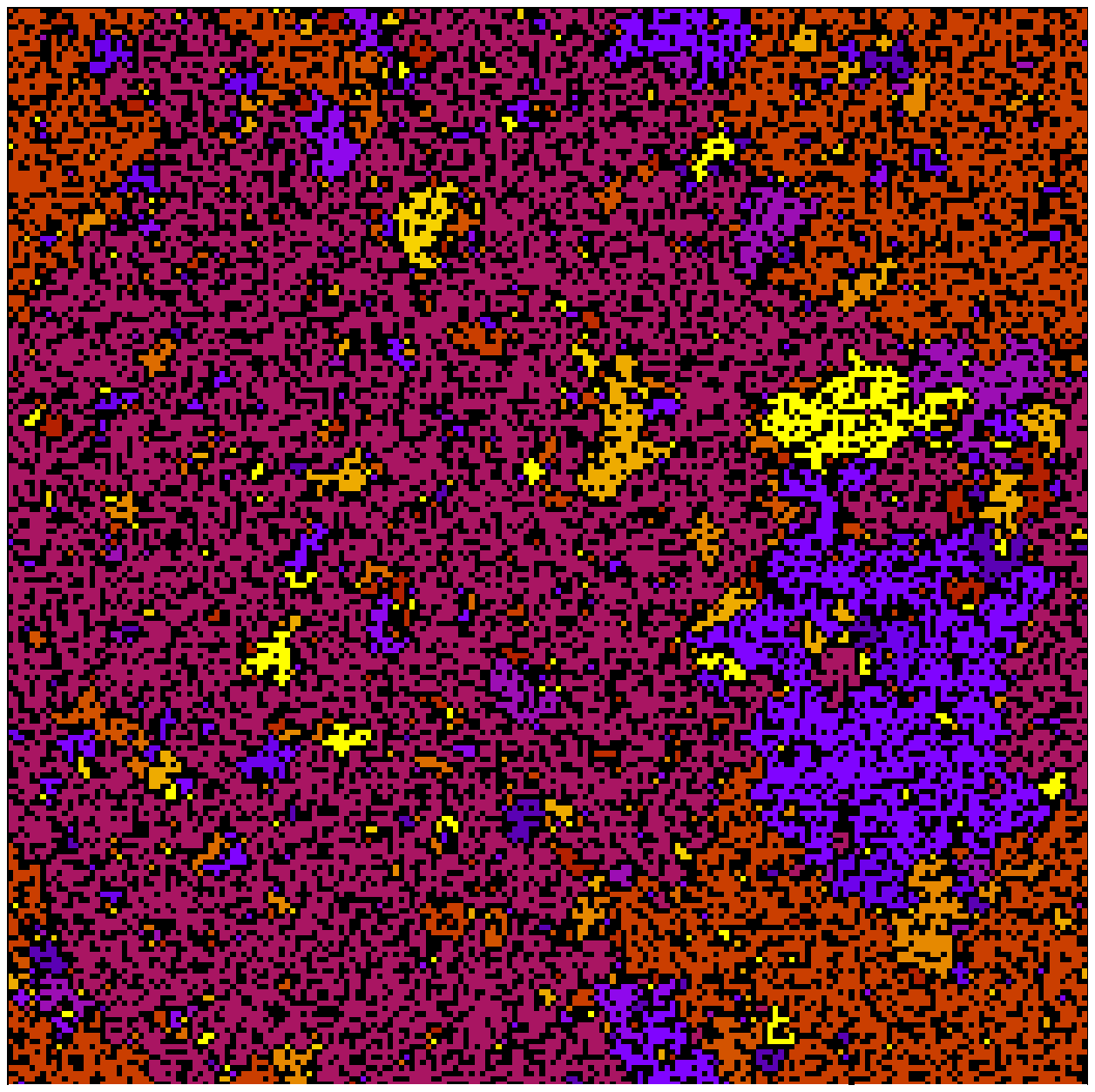}
	\hskip -3cm
	\includegraphics[width=0.93\columnwidth]{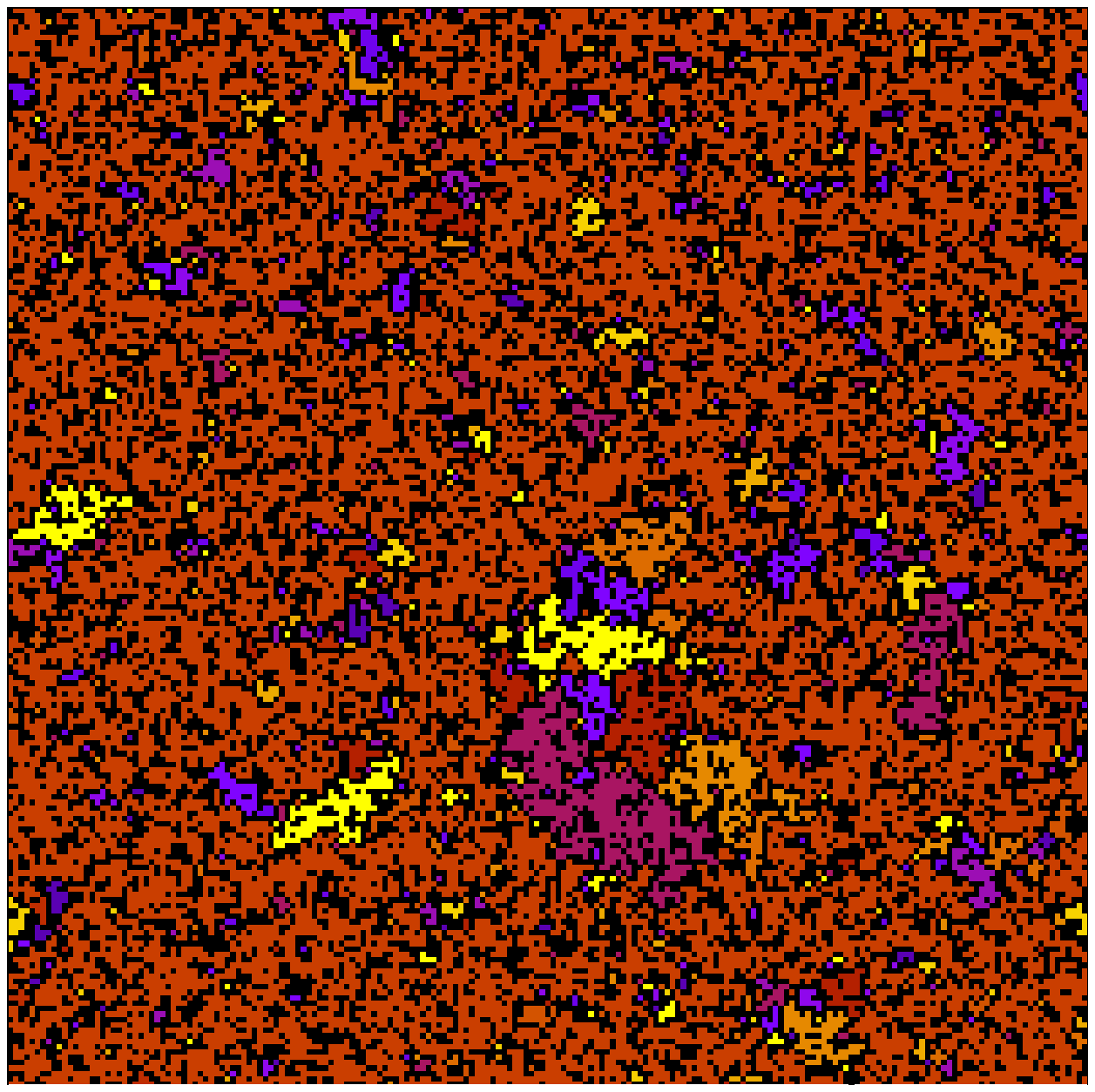}
	\hskip -3cm
	\includegraphics[width=0.93\columnwidth]{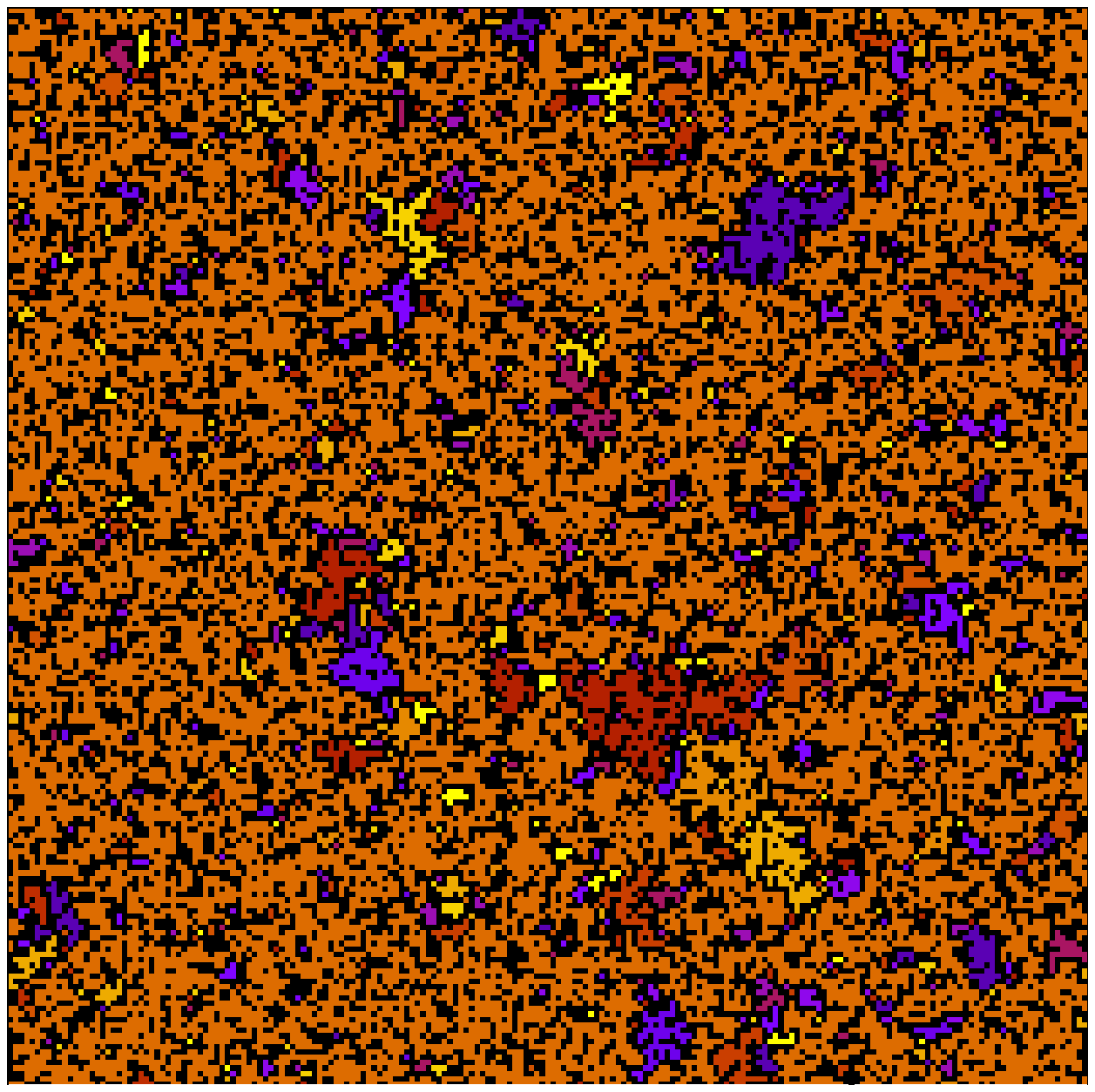}
	\vskip -3cm
\caption{Typical cluster structures of RCs on a $200 \times 200$ square lattice at an occupation probability, $x=0.594(1)$, slightly above the site-percolation threshold. Left panel: uncorrelated percolation, corresponding to the structure during relaxation with $n=1$. Middle panel: during induction with $p=0.9$ and $n=2$. Right panel: during induction with $p=0.9$ and $n=3$. Sites with the same colour represent connected clusters of closed RCs.\label{fig_3}}
\end{figure*}

During induction, open RCs are also closed through exciton wandering, which is possible if the exciton hops from a nearest neighbor closed site. In this way, correlations between nearest-neighbor sites are created.  These are, however, neglected in the LMF approximation presented above. To illustrate the bunching effect of exciton wandering we have calculated typical cluster structures of the RCs in our model on the square lattice at an occupation probability, $x=0.594(1)$, slightly above the site-percolation threshold ($x_{\textit{perc}}=0.5927460$\cite{site_perc,site_perc1}) with $n=1,2,$ and $3$; these are presented in Fig. \ref{fig_3}. For $n=1$, which corresponds to uncorrelated percolation and represents the state of the system during relaxation, the formation of a giant \textit{fractal} cluster is visible. For $n=2$ and $n=3$, which illustrate the state of the system during induction the giant cluster is visingly \textit{compact}. This is explained by the effect of exciton bunching, which causes a decrease in the critical percolation threshold so that the system is in the super-critical phase for the given value of $x$.

\begin{figure}[ht]
\includegraphics[width=1.0\columnwidth]{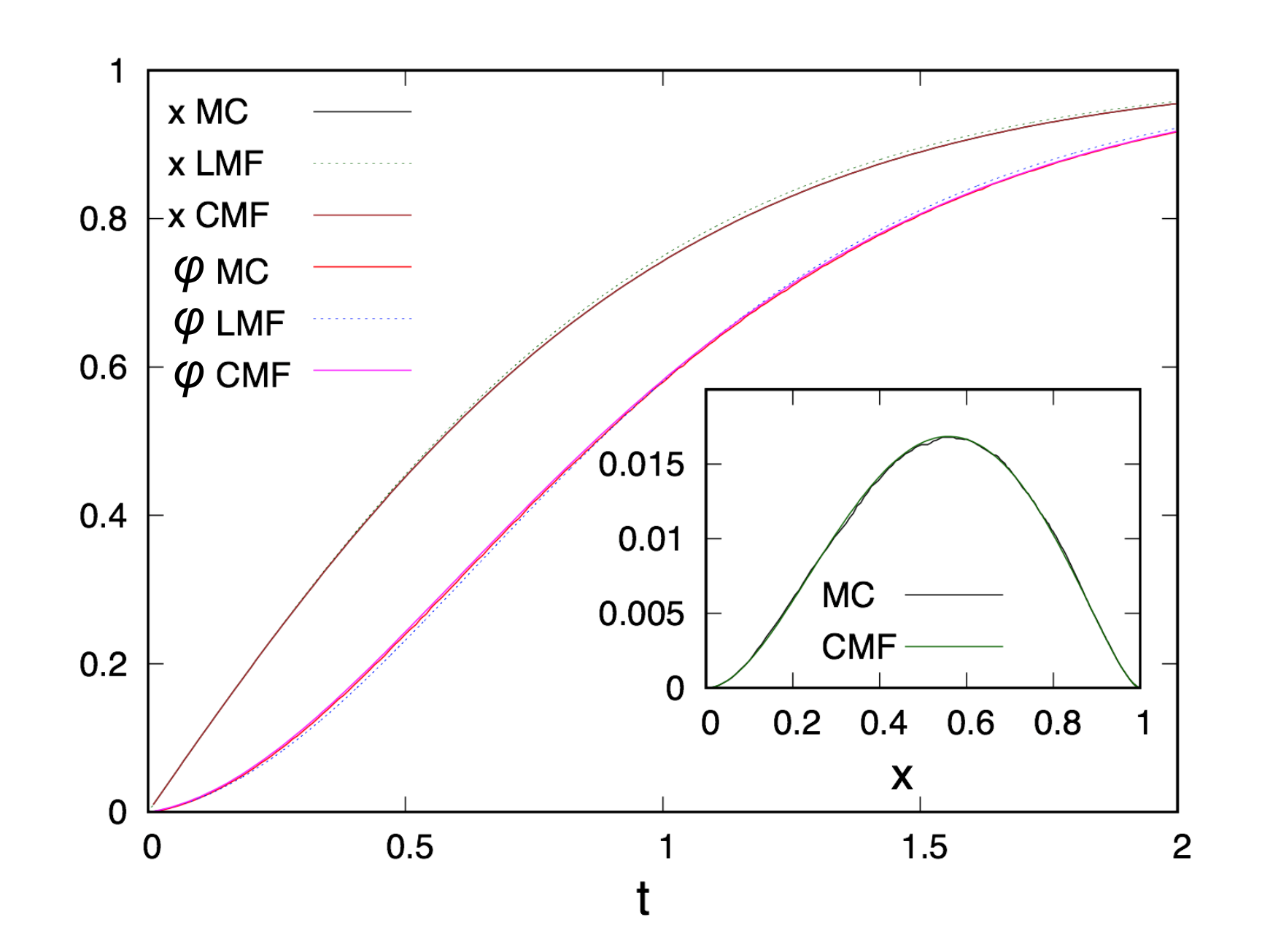}
\caption{Dynamics of the order-parameter, $x(t)$, and the fluorescence yield, $\varphi(t)$, calculated for a hopping probability $p=0.9$ and for $n=2$. Results of the LMF and CMF approaches are compared with MC simulations during induction. Inset: Connected nearest-neighbour correlation function $\tilde{x}_2=x_2-x^2$ as a function of $x$ for $p=0.9$ and for $n=2$. The CMF calculations perfectly overlap with the results of MC simulations.
\label{fig_4}}	
\end{figure}

We have calculated the nearest-neighbour correlation function, $x_2=\langle \sigma_{i_1}\sigma_{i_2}\rangle$ through MC simulations and compared the results with its uncorrelated value: $\langle \sigma_{i_1}\rangle\langle\sigma_{i_2}\rangle=x^2$. The difference, the connected correlation function $\tilde{x}_2=x_2-x^2$ is shown in the inset of Fig.\ref{fig_4} as a function of $x$, which is certainly not negligible, for $x \le 0.6$ their relative weight is about $10\%$. We have checked that a similar trend is present for larger values of $n=3$ and $4$, and that the correlations for nearest neighbors are larger than those between more remote sites (i.e. those having a distance of two or three lattice units).

Based on this observation we introduce the cluster mean-field approach, in which the bunching of closed RCs is taken into account through one more parameter, the nearest-neighbour correlation function, $x_2$. To obtain this, we solve the dynamics of a two-site cluster and the correlations which involve more sites are expressed in terms of two-site and one-site functions. For example the three-site function in this approach is given by:
\be
\langle \sigma_{i_1}\sigma_{i_2}\sigma_{i_3}\rangle \approx \frac{\langle \sigma_{i_1}\sigma_{i_2}\rangle\langle \sigma_{i_2}\sigma_{i_3}\rangle}{\langle \sigma_{i_2}\rangle}\;,
\label{two_appr}
\ee
while the general result is given in Eq.(\ref{G_k_CMF}). Having the analytical results in Eqs.(\ref{app:u}) and (\ref{app:q}) we have calculated the connected nearest-neighbour correlation function, $\tilde{x}_2=x_2-x^2$ on the square lattice, which is plotted for $p=0.9$ as a function of $x$ in the inset of Fig. \ref{fig_4}. Comparing it with the numerical values, obtained by MC simulations during induction an almost perfect agreement obtained.

We have also calculated the time-dependence of the order-parameter, $x(t)$, and that of the fluorescence yield $\varphi(t)$, in the CMF approach, the results are shown in Fig. \ref{fig_4} together with those calculated by the LMF approach as well as with MC simulations during induction.

It is seen in this figure that $x(t)>\varphi(t)$, which is due to exciton wandering. The results of CMF perfectly fit the MC simulations, at least within the numerical accuracy of the latter method. On the contrary the results of the LMF methods show small, but non-negligible differences. The LMF results overestimate $x(t)$ in particular for large $t$. For the fluorescence yield the LMF approach underestimates it at small $t$, but overestimates it for large $t$.

Finally, we calculate the relation between the fluorescence yield $\varphi$ and the fraction of closed RCs, $x$, and the results for the square lattice are presented in Fig.\ref{fig_1} for $n=2$ (main panel) and $n=3$ (inset), at a hopping probability $p=0.9$. Here we have made the calculations both during relaxation, when the LMF results are compared with MC simulations and with the (best fit of) the Joliot theory, and during induction, when the CMF results are compared with MC simulations and with the (best fit of) the Joliot theory. For $n=3$ the fluorescence yield from Eq.(\ref{F}) is given by:
\be
\varphi=(1-p)x+(1-p)px_2+p^2 \left(\tfrac{1}{4}x_2+\tfrac{3}{4}x_3\right)\;,
\ee
where the three-site function is written as $P_{\bullet,\bullet,\bullet} \equiv x_3 \approx (x_2)^2/x$, in agreement with Eq.(\ref{two_appr}).

As seen in Fig. \ref{fig_6} the analytical calculations agree very accurately with the MC simulations, both during relaxation (in which case the uncorrelated structure of the RCs perfectly fit with the similar assumptions of the LMF approach) and during induction (in which case the bunching effect of the closed RCs are well modelled in the CMF approach). In the curves there is a hysteresis, at the same value of $x$ the fluorescence yield is larger during induction, than during relaxation, which is due to the bunching effect. The hysteresis increases with larger value of $n$, since the bunching effect is also larger in this case. Concerning the Joliot theory the curve with $p=0.9$ fits only the starting part of the curves for small $x$, but deviates considerably for larger $x$ values. We can have an overall better description, if we set $p$ as a free parameter. Then, by using different best fit parameters during induction and relaxation the agreement with the measured curves becomes better, although still far less satisfactory, than the LMF and the CMF results.

\section{Discussion}
\label{sec:discussion}

The transfer of the excitons to the RC is extremely efficient as almost
every photon of the absorbed light in the antenna is used by the RC\cite{Wraight_and_Clayton_1974}.
The extreme efficiency of light utilization
supports the assumptions used above: the exciton is trapped or reflected
by collisions with open or closed RCs, respectively and the redirected
exciton can visit several other RCs. Indeed, the RC of the \textit{cycA} mutant
acts accordingly: the open state PQ\textsubscript{A} is a perfect trap
and the closed state
P\textsuperscript{+}Q\textsubscript{A}\textsuperscript{--} is a perfect
reflector of the incoming excitons. In this variant, the oxidized dimer
P\textsuperscript{+} of the closed RC does not allow any additional
charge separations including the short lived
P\textsuperscript{+}BPheo\textsuperscript{--} radical pair. There is no
known exciton-radical pair equilibrium, whose "reverse reaction" would
be required to apply standard (homogeneous kinetic) models\cite{Lavergne_and_Trissl_1995}.
This introduces the phenomenological concept of imperfect traps for open
($0.25 \pm 0.05$) and closed ($0.40 \pm 0.05$) RCs (\emph{R. rubrum}) to
describe the characteristics (e.g. the initial and maximum levels) of
the fluorescence induction\cite{Timpmann_et_al_1993}. The probability of
redirection of the exciton from the open RC to the antenna has been
estimated between $5-30\%$ in various purple bacteria\cite{Lavergne_and_Trissl_1995}.
In contrast, the exciton walking approach does not need
this \emph{ad hoc} assumption, instead, it considers the RC as a perfect
trap (for photochemistry) or reflector (for migration) of the excitons.

A long-standing question is how the efficiency of the bacterial antenna can be so high at ambient temperature given that it is a partly disordered biological system. The structural
data from atomic force\cite{Scheuring_et_al_2005,Liu_and_Scheuring_2013}
and cryo-electron microscopy\cite{Xin_et_al_2018} and functional results
from two-dimensional electron spectroscopy and related calculations\cite{Kramer_and_Rodriguez_2017} clearly demonstrate the close packing of
the BChl complexes and the strong coupling, respectively, which are the
necessities for exciton formation. One can ask whether the funnelling of the excitation energy to the RC occurs through random hops or straight walks of the exciton? The interaction among the chromophores
within the PSU can be so high that even the signs of quantum coherence
may appear\cite{Panitchayangkoon_et_al_2010,Ishizaki_and_Fleming_2012}. Currently, the temptation is large to attribute the quantum
coherence observed in the antenna system of photosynthetic organisms to
be similar to that in quantum computers\cite{Ball_2018}. However, the energetic
coupling among the PSUs is not so large as among the chromophors within
the PSU. The smaller connectivity permits a random walk rather than a
direct walk of the excitons to the nearest open RC. This is why we
pictured the movement of the excition as a random (incoherent) hopping
process. The excitation at an arbitrary site of the antenna does not find an optimal route to the nearest open RC but has to waste time through random hopping.

The random walk approach applied in this study drops two essential
simplifications which limit the validity of the Joliot theory. 1) The
exciton redirected from a closed RC can visit any RC (independent of
their relative locations) with probability \emph{p} (Joliot parameter)
or with connectivity parameter \emph{J} = \emph{p}/(1--\emph{p}). This
is a disputed assumption of the Joliot model as the rate of energy
transfer between donor and acceptor chromophores has strong
distance-dependence (see the inverse power 6 dependence of the rate
constant via dipole-dipole interaction in the Förster mechanism). The
transfer (hop) to the neighboring RC is more probable than to a distant
RC. The real motion of the exciton is adequately treated by a random
walk on the network of the RCs as used in our model. 2) The distribution
of the closed RC is taken randomly at any moment of the kinetics.
However, this assumption is true during the relaxation only and fails
during the induction. On the one hand, the fraction of closed RC in the
relaxation process is controlled by the chemical re-reduction of
P\textsuperscript{+} and the distribution remains always random during
the decay. On the other hand, when the RCs are closing progressively
under a continuous excitation (induction), the distribution of the
closed RCs will not be random due to bunching effects: an open RC has
higher chances to become closed when its neighbors are already closed.
The distribution will differ from the Poisson distribution, and will
depend on the degree of saturation (i.e. on the time). The simultaneous
fluorescence and absorption change kinetics observed both during induction
and during relaxation indicate clearly the limits of the standard theory
and the experimental manifestation of the bunching effect (Figs. \ref{fig_02} and
\ref{fig_03}). The concavities of the $\varphi(x)$ curves were different: it
was smaller during induction than during relaxation and the difference
(hysteresis) seemed to be dependent on the physiological state (age) of
the bacteria.

Here we used an exciton migration model in which the possible pathways
of the exciton were represented with different approximations. In the
homogeneous kinetic model the exciton could hop to any RC irrespective
of its distance and position and the PSU dynamics was treated in the
(one-site) mean-field level. In the LMF approach the exciton hoping was
restricted to nearest neighbour RCs and the same closed RC could be
visited several times, but the PSU dynamics was still in the mean-field
level. Finally, in the CMF approach, while the exciton wandering respects the local topology of the RCs, the PSU dynamics were treated at the (two-site) cluster level. In this way bunching of closed RCs during
induction was taken into account and the experimentally observed
hysteresis could be successfully explained. The basic ingredients and
approximations which were used in the different approaches are
summarised in Table \ref{table_sum}.
%
\begin{table}[ht]
\begin{tabular}{|c|c|c|c|}
\hline
 & Joliot-theory & LMF & CMF \\
 \hline
density of closed RCs& yes & yes & yes \\
 \hline
dynamics of $x$  & yes & yes & yes \\
\hline
lattice topology & no & yes & yes \\
\hline
multiple visits of sites & no & yes & yes\\
\hline
bunching of closed RCs& no & no &yes\\
\hline
hysteresis & no & no &yes\\
 \hline
\end{tabular}
\caption{\label{table_sum} Basic ingredients involved in the different approaches.}
\end{table}

\begin{figure}[ht]
\hskip 0cm
\includegraphics[width=1.15\columnwidth]{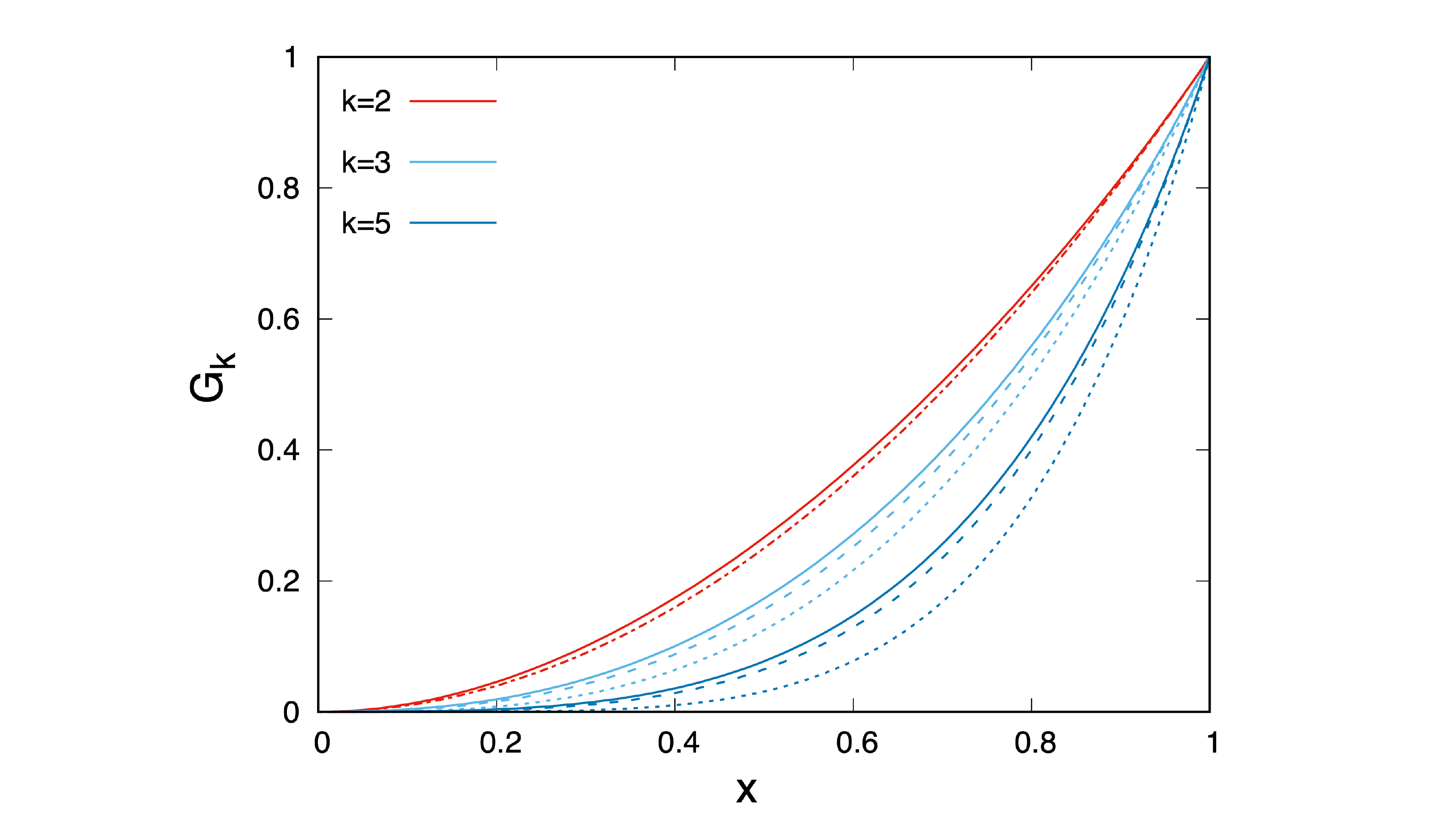}
\vskip 0cm
\caption{$k$-site correlations vs. fraction of closed RCs calculated in the different approaches at a hopping probability $p=0.9$. Dotted line: Joliot-theory, dashed line: LMF approach, full line: CMF approach. 
\label{fig_5}}	
\end{figure}

The multi-site correlation functions, $G_k=\langle \sigma_{i_1}\sigma_{i_2} \dots \sigma_{i_k}\rangle$, were introduced as
fundamental quantities of the theoretical treatment (see Eq.(\ref{m-corr})). $G_k$ is the fraction of such $k$-step random walks (modelling exciton wandering), which visit (nearest neighbour) closed RCs. The approximate representation of $G_k$ in the different approaches is described in details in Sec. \ref{sec:mathematical_methods}. The $x$ dependence of $G_k$ is illustrated in Fig.\ref{fig_5} for different values of $k$ at a hopping probability $p=0.9$. As a general rule $G_k$ is larger if the exciton finds more closed RCs at the nearby steps. Therefore $G_k$ is a monotonously increasing function of $x$. For a given value of $k$, $G_k$ is the smallest in the Joliot theory, in which multiple visits of the same closed RC don't take place. Comparing the results from the CMF and LMF approaches, the former is somewhat larger due to the bunching effect. Based on the multi-site correlations, some essential quantities can be calculated, such as the absorption cross-section of the RC (due to the presence of closed RCs in the neighborhood) and the average number of exciton steps during migration. 

\begin{figure}[ht]
\vskip 0cm
\includegraphics[width=1.0\columnwidth]{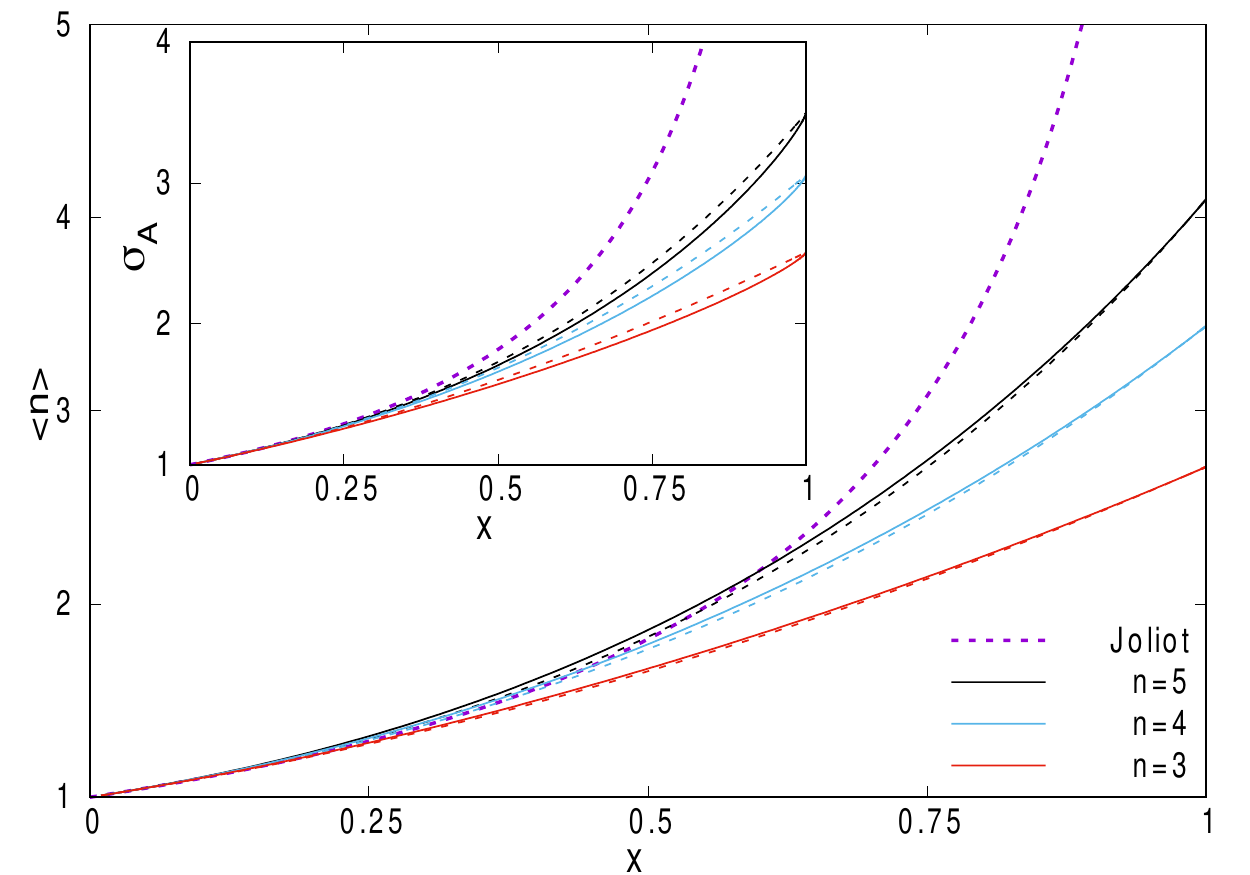}
\vskip 0cm
\caption{Average number of exciton steps as a function of the fraction of closed RCs at a hopping probability $p=0.9$ calculated with the CMF approach (full line) and with the LMF approach (dashed line), for different maximal number of steps, $n$. With dotted line result of the Joliot-theory ($n \to \infty$) is presented. Inset: absorption cross section as a function of the fraction of closed RCs.
\label{fig_6}}	
\end{figure}

The absorption cross section, $\sigma_A$, calculated in the different approaches is shown in the inset of Fig. \ref{fig_6} at a hopping probability $p=0.9$. It is a monotonously increasing function of \emph{x}, as
more and more RCs will be closed in the vicinity. It is also
increasing with \emph{n}, when more exciton steps can be made. Since in
the Joliot theory \emph{n} is unlimited, the corresponding absorption
cross section is much larger, than those for finite values of \emph{n}.
Having the same value of \emph{n}, $\sigma_A$ is somewhat larger during relaxation (which corresponds to the LMF approach), than during induction (described with the CMF method). Due to bunching the exciton stays longer on closed RCs in the latter process, and has smaller probability to reach an open one.

The average number of exciton steps, $\langle n \rangle$, calculated in the different approaches is shown in the main figure of Fig. \ref{fig_6} at a hopping probability $p=0.9$.
It is seen, that $\langle n \rangle$ is a monotonously increasing function of $x$ and has its maximum at $x=1$: $\langle n \rangle =\dfrac{1-p^n}{1-p}$. The general shape of the curves is similar to that of the absorption cross section in the inset of Fig. \ref{fig_6}, with the difference, that for a given $n$, its average value is larger during induction (CMF method), than during relaxation (LMF approach). Indeed, due to bunching, the exciton finds closed RCs with higher probability in the former process.

Any changes of the physiological state reflect adaptation of the bacterium to the variable environmental conditions with the goal of establishing a fine balance against several requirements. The changes of
the light intensity result in changes of the antenna organization and
exciton migration\cite{Niederman_2013,Niederman_2016}. Under light-limiting
conditions, the light must be collected with higher efficiency by
increase of the LH2 antenna size and of the connectivity of the PSUs. At
higher light intensity, photobleaching becomes the bottleneck. The
fraction of closed RCs comes closer to saturation resulting in the
increase of the absorption cross section of the RCs (inset of Fig. \ref{fig_6}) and in the
number of steps (and thus the lifetime) of the migrating excitons (Fig.
\ref{fig_6}). These effects enhance the probability of BChl triplet formation and
make the bacterium more vulnerable to photooxidation. The response of
the cell to these conditions is the reduction of the energetic coupling
of the PSUs by loosing the antenna structure including setting spacers
between the LH complexes. The loose-fitting core complex may facilitate
the diffusion of the quinone that shuttles electrons and protons between
RC and cyt \emph{bc}\textsubscript{1} complex\cite{Comayras_et_al_2005,Olsen_et_al_2017}. Tentatively, we assigned the observed dependence of
hysteresis on the duration of the cultivation of the bacteria to changes
of the membrane packing. Further work is required to confirm and
understand the effect.

The intactness of the cells had special importance in this study as all earlier works referred to chromatophores. While the optical signal was disturbed by light scattering due to the larger sizes of the cells, we were able to diminish its effect and obtain optical signals with quality close to those obtained from chromatophores. By use of whole cells instead of chromatophors prepared by invasive biophysical and biochemical methods, we could preserve the physiological state of the bacterium.
 
The observed hysteresis (bunching effect) may include unexpected and interesting manifestation of a memory function. That is, the induction process is influenced by a sort of memory of the way the given state was prepared. Near-neighbour correlations are induced between the closed RCs, thus their distribution is not fully random. This correlation can be taken as the expression of the memory of the state. On the contrary, the relaxation process is controlled by spontaneous re-reduction of P\textsuperscript{+} (via charge recombination or external electron donor), thus the distribution of the closed RC-s is completely random and uncorrelated. No “memory function” can be introduced.

For a fixed fraction of closed PSUs (\emph{x}), it is interesting to see
how the fluorescence yield depends on the connectivity of the units
(Figs. \ref{fig_04} and \ref{fig_05}). As expected, the energetic coupling (\emph{p})
decreases the fluorescence yield but the decline depends on the pattern
of the distribution of the closed RCs (hysteresis): it is smaller in
induction than during relaxation under otherwise identical conditions. The correlated clusters of closed centers during the induction phase results in larger fluorescence than the uncorrelated clusters during the relaxation phase. In the case of too much excitation (including not only high exciton density but a large fraction of closed RCs, as well), the
fluorescence can be considered as a valve function of energy dissipation
that is useful for photoprotection \cite{Sipka_and_Maroti_2018}. The
observed phenomenon of hysteresis may reveal new aspects of the
competition between harvest and dissipation of light energy and can
serve as a fine tuning mechanism of the light utilization in the
antenna.\\

\section{Methods}
\label{sec:methods}

\subsection{Materials and Experimental Methods}
\label{sec:materials}

\emph{Cytochrome c less bacterial mutant strain and chemicals}: The
\textit{cycA} mutant was constructed from a wild type strain of purple
nonsulfur photosynthetic bacterium \emph{Rhodobacter (Rba.) sphaeroides} as
described earlier\cite{Sipka_et_al_2018}. Strain JS2293$\Delta$, containing an
in-frame deletion of the \textit{cycA} gene encoding cytochrome
\emph{c}\textsubscript{2} in \emph{Rba. sphaeroides}, was genetically
constructed essentially as described previously\cite{Chi_et_al_2015}.
\emph{Escherichia coli} strains were grown at 37\textsuperscript{o}C
in LB medium\cite{Sambrook_et_al_1989} supplemented with antibiotics when appropriate;
kanamycin (50 $\mu$g mL\textsuperscript{--1}) and ampicillin (100 $\mu$g
mL\textsuperscript{--1}). \emph{Rba. sphaeroides} strains were grown
aerobically at 30\textsuperscript{o}C in YCC medium\cite{Sistrom_1977} supplemented when appropriate with
kanamycin (50 $\mu$g mL\textsuperscript{--1}). Conjugal
transfer of strains from \emph{E. coli} to \emph{Rba. sphaeroides} was
performed as described previously, and counter-selection against S17-1
donors was achieved by addition of tellurite (100 $\mu$g
mL\textsuperscript{--1})\cite{Donohue_and_Kaplan_1991}. The cyt
\emph{c}\textsubscript{2} mutant bacteria were cultivated in a half
filled Erlenmeyer flask plugged by rolls of cotton wool (semiaerobic
conditions) on a shaking plate in the dark. The increase of
concentration of bacteria saturated three days after the inoculation and
samples could be taken in different phases of the bacterial growth (see\cite{Kis_et_al_2014}).

To inhibit the interquinone electron transfer after flash excitation,
terbutryne was used in 120 $\mu$M concentration. The herbicide terbutryn has
proved to be highly efficient even in whole cells of bacteria to block
the Q\textsubscript{A}\textsuperscript{--}Q\textsubscript{B}$\to$
Q\textsubscript{A}Q\textsubscript{B}\textsuperscript{--} electron
transfer in the acceptor quinone complex of the RC by competition with
the pool quinones for the same secondary (Q\textsubscript{B}) binding
site.

\emph{Optical measurements}

The transient changes of absorption and fluorescence of the intact cells were generated by high power (2 W) laser diodes using variable flash durations. Flashes that were approximately 1 ms in duration were energetically sufficient to cause the gradual closure of all of the mutant RCs. Red (wavelength 804 nm) or blue (wavelength 450 nm) laser
diodes were applied to excite the BChl dimer P of the RC directly via
BChls of the LH2 or indirectly through accessory pigments (e.g.
carotenoids). The different excitation modes delivered very similar
results.

\emph{Absorption}: Because the light-induced oxidation of the RC dimer induces an electrochromic shift in the absorption band of the nearby monomeric BChl\cite{Bina_et_al_2010}, the kinetic status of the oxidized
dimer (P\textsuperscript{+}) was tracked by measurement of the absorption change at 790 nm. The weak measuring beam was chopped for long periods of time by a mechanical shutter to avoid the excitation of the sample during both induction and relaxation. As the magnitude of the absorption change proved to be small
($\Delta$\emph{A} $\sim$ 1 mOD), the absorption kinetics were acquired
as averages of several (up to 64) scans to reduce the statistical error.
The rate of repetition of the flashes had to be fitted to the complete
relaxation of the
P\textsuperscript{+}Q\textsubscript{A}\textsuperscript{--} charge
separated state ($\sim$ 20 s). More (128) scans were needed to measure the absorption changes during relaxation where the small signal-to-noise was overlapped by the slow drift of the baseline in the prolonged time scale of about 10 seconds.

\emph{Fluorescence}: The home-built experimental set-up (BChl
fluorometer) and the data processing of fluorescence of intact cells
have been described in detail\cite{Kocsis_et_al_2010}.
The fluorescence (wavelength centered at 900 nm) from sample in $3 \times 3$ mm
quartz cuvette was measured during the excitation during induction mode and
detected by testing flashes in relaxation mode. The fluorescence quanta
emitted in the direction perpendicular to the actinic light beam were
detected by a near-infrared-sensitive, large-area (diameter 10 mm), and
high-gain Si-avalanche photodiode (APD; model 394-70-72-581; Advanced
Photonix Inc., USA, working resistance 1.5 k$\Omega$). A long pass filter (RG
850, Schott) was used to protect the detector from scattered light of
the laser and to cut off fluorescence emission from the other pigments
than BChl and the base plate. Solutions of extracted BChl or IR-806 dye (Sigma) served as references for fluorescence yield measurements and to correct for any deviations from the step function (rectangular shape) and for large-scale fluctuation of the laser diode excitation. The reference
signal was adjusted to the same intensity as that of the fluorescence to
avoid the possible artefact coming from the nonlinearity of the response
of the detector at high light intensity.

The extreme values of the induction kinetics of fluorescence (\emph{F})
were determined experimentally as follows. The constant part of the
fluorescence rise (\emph{F}\textsubscript{0}) was obtained by
interception of the initial data (approximated by straight line) and the
vertical axis at \emph{t} = 0 and the maximum fluorescence
(\emph{F}\textsubscript{max}) by the saturating value of the induction.
The normalized variable fluorescence was derived as $\varphi$ =
(\emph{F}-\emph{F}\textsubscript{0})/(\emph{F}\textsubscript{max}-\emph{F}\textsubscript{0}).
The fluorescence of the sample during relaxation was probed by a couple
($\sim$15) of intense but short (5 $\mu$s) laser flashes. The
non-exciting character of the testing flashes was checked before each
experiment.

All measurements were performed at room temperature (20--25
\textsuperscript{o}C).

\subsection{Mathematical Methods}
\label{sec:mathematical_methods}

The multi-site correlation function $G_k = \langle \sigma_{i_1}\sigma_{i_2} \dots \sigma_{i_k}\rangle$, is the fraction of such $k$-step random walks which visit (nearest neighbour) closed RCs, see in Eq.(\ref{m-corr}).
In the \textit{Joliot theory} multiple visits of the same site is excluded and the multi-site correlations are approximated as products of one-site functions, $P_{\bullet}=x$:
\be
G_k \to P_{\bullet} \ast P_{\bullet} \ast \dots \ast \mathrel{\overset{\makebox[0pt]{\mbox{\normalfont\tiny\sffamily  k }}}{P_{\bullet}}}=x^k\;.
\label{G_k_LMF}
\ee

\textit{Lattice mean-field approach}

In the LMF approach the exciton hops on nearest-neighbour lattice sites, multiple visits of the same site is allowed and the local topology of the lattice is encoded in the weights, $c_j^{(k)}$, $j=2,3,\dots,k$. Reduced multi-site correlations are approximated as product of one-site functions:
\beqn
G_k &\to& c_2^{(k)}P_{\bullet}\ast P_{\bullet} + c_3^{(k)}P_{\bullet}\ast P_{\bullet}\ast P_{\bullet} + \dots +\nonumber \\
&+& c_k^{(k)}P_{\bullet}\ast P_{\bullet}\ast \dots \ast \mathrel{\overset{\makebox[0pt]{\mbox{\normalfont\tiny\sffamily k}}}{P_{\bullet}}}=\sum_{j=2}^k c_j^{(k)}x^j\;.
\eeqn
%
%
\begin{table}[h!]
\begin{tabular}{|c|r|r|r|r|r|r|r|r|r|}
\hline
 k&2 & 3 & 4 & 5 & 6 & 7 & 8 & 9 & 10\\
 \hline
j=2  & 1 & 1 & 1 & 1 & 1 & 1 & 1 & 1 & 1\\
 3&  & 3 & 6 & 12 & 18 & 30 & 42 & 66 &90\\
 4&  &  & 9 & 26 & 72 & 161 & 338 & 690 &1317\\
 5&  &  &  &  25&  94& 319 & 890 & 2335 &5668\\
 6&  &  &  &  & 71 & 318 & 1256 & 4066 & 12325\\
 7&  &  &  &  &  & 195 & 1026 & 4515 &16434\\
 8&  &  &  &  &  &  & 543 & 3232 &15692\\
 9&  &  &  &  &  &  &  & 1479 &9942\\
 10&  &  &  &  &  &  &  &  &4067\\
 \hline
\end{tabular}
\caption{\label{table_c_i_n} List of $C_j^{(k)}/4$, where $C_j^{(k)}$ is the number of $(k \ge 2)$-step random walks, which have visited $2 \le j \le k$ different sites on the square lattice, where the walker arrives to the lattice at the first step, see text. These parameters appear as the weights of the polynomials in Eqs.(\ref{m-corr-rel}) and (\ref{c_j_k}).}
\end{table}
%
The weights are calculated through random walk statistics using the parameterisation: 
\be
c_j^{(k)}=C_j^{(k)}/z^{k-2}\;,
\label{c_j_k}
\ee
where $C_j^{(k)}/z$ is the number of $(k \ge 2)$-step random walks, which have visited $2 \le j \le k$ different sites, where the walker arrives to the lattice at the first step. For the square lattice with $z=4$ the first few terms of $C_j^{(k)}$ are given in Table \ref{table_c_i_n}.

The dynamics of $x$ is calculated from Eq.(\ref{dx/dt}) and with separation of the variables it is given by:
\be
\int_0^x \frac{\textrm{d} x'}{1-\varphi(x')} = t\;.
\label{int}
\ee
Here the denominator is a polynomial of $x'$, and thus can, in principle, be integrated for all values of $n$. For the first three values of $n$ these are given by:
\beqn
&&x(t)=1-\exp(-t),\quad n=1\;, \label{n=1}\\
&&x(t)=\frac{\exp[(1+p)t]-1}{\exp[(1+p)t]+p},\quad n=2\;,\label{n=2}
\eeqn
\beqn
&&t= \frac{(1+1.5p)/\sqrt{2}}{(1+p+3/4p^2)}\left[\arctan\left(\frac{1.5px +1}{\sqrt{2}}\right)-\arctan\frac{1}{\sqrt{2}}\right]\nonumber\\
&&-\frac{1}{2(1+p+3/4p^2)}\ln\frac{|1-x|^2}{|1+px +3/4(px)^2 |},\quad n=3\;.\label{n=3}
\eeqn
 The LMF approach is expected to be appropriate in the relaxation process.

\textit{Cluster mean-field approach:}

In the CMF approach the exciton hops on nearest-neighbour lattice sites, multiple visits of the same site is allowed and
the local topology of the lattice is encoded in the weights, $c_j^{(k)}$, $j=2,3,\dots,k$.
Multi-site correlations are expressed in terms of two-site ($P_{\bullet,\bullet}=x_2$) and one-site functions:
\beqn
&G_k& \to c_2^{(k)}P_{\bullet,\bullet} + c_3^{(k)}\frac{P_{\bullet,\bullet} \ast P_{\bullet,\bullet}}{P_{\bullet}} + \dots +\nonumber \\
&+&c_k^{(k)}\frac{P_{\bullet,\bullet} \ast P_{\bullet,\bullet} \ast \dots \ast \mathrel{\overset{\makebox[0pt]{\mbox{\normalfont\tiny\sffamily k-1}}}{P_{\bullet,\bullet}}}}{P_{\bullet} \ast P_{\bullet} \ast  \dots  \ast \mathrel{\overset{\makebox[0pt]{\mbox{\normalfont\tiny\sffamily k-2}}}{P_{\bullet}}}}=
 \sum_{j=2}^k c_j^{(k)}\frac{x_2^{j-1}}{x^{j-2}}\;.
 \label{G_k_CMF}
\eeqn
The basic correlations in the CMF approach are calculated for a two-site cluster, in which case the occupation probabilities of the different configurations are given by: $P_{\circ,\circ}$, $P_{\circ,\bullet}$, $P_{\bullet,\circ}$ and $P_{\bullet,\bullet}$. Here $P_{\circ,\bullet}=P_{\bullet,\circ}$ due to symmetry and $P_{\circ,\circ}+2P_{\circ,\bullet}+P_{\bullet,\bullet}=1$, due to normalisation. Thus we have two independent parameters: $x=P_{\circ,\bullet}+P_{\bullet,\bullet}$ and $x_2=P_{\bullet,\bullet}$, so that $P_{\circ,\bullet}=x-x_2$ and $P_{\circ,\circ}=1-2x+x_2$.

The time-dependence of the one-site function is given by:
\be
-\frac{\textrm{d} P_{\circ}}{\textrm{d} t}=P_{\circ}+p'zP_{\circ,\bullet}\;,
\label{dx/dt_1}
\ee
where $p'$ is the hopping probability from the $\newmoon$ site of the $\fullmoon \newmoon$ cluster, having $z$ different orientations in the lattice. This quantity can also be calculated from Eq.(\ref{dx/dt}), when the r.h.s. of Eq.(\ref{dx/dt_1}) is given by: $1-F=1-(1-p)x-px_2=1-x+p(x-x_2)=P_{\circ}+pP_{\circ,\bullet}$, thus we obtain a relation between the hopping probabilities: $p=p'z$. The time-derivative of the two-point function, $P_{\circ,\circ}$, involves the three-point function, $P_{\circ,\circ,\bullet}$, in which the $\newmoon$ site can be in $2(z-1)$ relative positions with respect to the $\fullmoon \fullmoon$ cluster:
\beqn
-\frac{\textrm{d} P_{\circ,\circ}}{\textrm{d} t}&=&2P_{\circ,\circ}+2p'(z-1)P_{\circ,\circ,\bullet} \nonumber\\
&=&2P_{\circ,\circ}+2p\frac{z-1}{z}\frac{P_{\circ,\circ}P_{\circ,\bullet}}{P_{\circ}}\;.
\label{dx/dt_2}
\eeqn
In the second equation the three-point function is approximated according to Eq.(\ref{two_appr}) and we use $p$ instead of $p'$.

To solve the coupled differential equations in Eqs.(\ref{dx/dt_1}) and (\ref{dx/dt_2}) first let us introduce the notations: $P_{\circ}=q$, $P_{\circ,\circ}=u$, so that $P_{\circ,\bullet}=q-u$, in terms of which the equations read as:
\beqn
-\frac{\textrm{d} \ln q}{\textrm{d} t}&=&1+p\frac{q-u}{q}\nonumber \\
-\frac{1}{2}\frac{\textrm{d} \ln u}{\textrm{d} t}&=&1+p\frac{z-1}{z}\frac{q-u}{q}\;.
\eeqn
Combining the two equations: 
\be
\frac{z-1}{z}\frac{\textrm{d} \ln q}{\textrm{d} t}-\frac{1}{2}\frac{\textrm{d} \ln u}{\textrm{d} t}=\frac{1}{z}\;,
\ee
from which we obtain the relation:
\be
u(t)=q^{2\tfrac{z-1}{z}}e^{-\tfrac{2}{z}t}\;,
\label{app:u}
\ee
with
\be
-\frac{\textrm{d} \ln q}{\textrm{d} t}=1+p-pq^{\tfrac{z-2}{z}}e^{-\tfrac{2}{z}t}\;.
\label{dq/dt}
\ee
\\
The solution of (\ref{dq/dt}) for $z=2$, which corresponds to the one-dimensional case, is given by:
\be
q(t)=\exp[-(1+p)t+p(1-e^{-t})],\quad z=2\;,
\ee
whereas for $z>2$ it is in the form:
\be
q(t)=\exp\left(\tfrac{2t}{z-2}\right)\left(\frac{p \tfrac{z-2}{z}+\exp\left[t\left(1+p\tfrac{z-2}{z}\right)\right]}{p \tfrac{z-2}{z} +1}\right)^{\tfrac{z}{2-z}},\quad z>2\;.
\label{app:q}
\ee
Then the basic correlation functions are given by: $x(t)=1-q(t)$ and $x_2(t)=1-2q(t)+u(t)$.

The CMF approach is expected to contain the basic physical background for the induction dynamics.

\textit{The photochemical utilization (absorption) of the exciton} can be realised after $k=1,2,\dots, n$ steps, provided the RC in the $k$-th step is open, but the RCs are closed in the previous $k-1$ steps. The sum of these contributions is given by:
\beqn
A&=&\langle 1-\sigma \rangle+p\langle \sigma_{i_1}(1-\sigma_{i_2})\rangle + p^2\langle \sigma_{i_1}\sigma_{i_2}(1-\sigma_{i_3})\rangle\nonumber\\
&+&\dots+p^{n-1}\langle \sigma_{i_1}\sigma_{i_2} \dots \sigma_{i_{n-1}}(1-\sigma_{i_n})\rangle=\nonumber\\
&=&\sum_{k=1}^{n}p^{k-1}(G_{k-1}-G_k) \;,
\label{A}
\eeqn
where $G_0=1$, $G_1=x$ and $A+\varphi=1$. Then the absorption cross section is defined as:
\be
\sigma_A=\frac{A}{P_{\circ}}=\frac{1-\varphi}{1-x}\;,
\ee
which in the Joliot theory with Eq.(\ref{Joliot_F}) is given by:
\be
\sigma_A=\dfrac{1}{1-px},\quad \textrm{Joliot theory}\;.
\label{sigma_A_Joliot}
\ee

\textit{The average number of steps}, $\langle n \rangle$, is given by:
\beqn
\langle n \rangle&=&(1-p)\sum_{k=1}^{n-1}kp^{k-1}G_k+np^n G_n\nonumber\\
&+&\sum_{k=1}^{n}kp^{k-1}(G_{k-1}-G_k)\nonumber\\
&=&\sum_{k=0}^{n-1}p^{k}G_k=1+\frac{p}{1-p}(\varphi-p^{n-1}G_n) \;.
\label{n_av}
\eeqn
Here in the first and in the second line the contributions from the steps ending with fluorescence and with absorption, respectively, are presented. In the Joliot theory $\langle n \rangle$ is given by:
\be
\langle n \rangle=\dfrac{1}{1-px},\quad \textrm{Joliot theory}\;,
\label{n_joliot}
\ee
which is just the absorption cross section in Eq.(\ref{sigma_A_Joliot}). It is interesting to note a relation with the absorption:
\be
A=\langle n \rangle-\frac{1}{p}(\langle n+1 \rangle -1)\;,
\ee
where $\langle n+1 \rangle$ is the average number of steps in such a process, in which the maximal number of exciton hops is $n+1$.

\begin{acknowledgments}
This work was supported by by EFOP-3.6.2-16-2017-0005 and by the National Research Fund under Grants No. K128989, No. K115959 and No. KKP-126749. I.A.K. was supported by the Domus Hungary Scholarship of the Hungarian Academy of Sciences. This publication was made possible through the support of a grant from the John Templeton Foundation. The opinions expressed in this publication are those of the authors and do not necessarily reflect the views of the John Templeton Foundation. J.L.S gratefully acknowledges support from the Bob and Roberta Blankenship endowment, the van Dyck Faculty Leave, and the Regan Faculty Leave.
\end{acknowledgments}

\textbf{Author contributions}

P.M.initiated the project and supervised all experiments. J.L.S.
engineered the cyt \emph{c}\textsubscript{2} less mutant of \emph{Rba.
sphaeroides}. M.K. cultivated the bacteria and performed the
experiments. I.A.K. and F.I. performed the exciton migration model and
MC simulation. P.M. and F.I wrote the manuscript.

\textbf{Competing interests}

The authors declare no competing interests.\\


\vskip 3cm
\section*{References:}
\begin{thebibliography}{99}

\bibitem{Mirkovic_et_al_2016}
Mirkovic, T., Ostroumov, E.E., Anna, J.M., van Grondelle, R., Govindjee
\& Scholes, G.D. Light Absorption and Energy Transfer in the Antenna
Complexes of Photosynthetic Organisms. \emph{Chem. Rev}.,
\textbf{117}(2), 249-293, (2016).

\bibitem{Maroti_and_Govindjee_2016}
Maróti, P. \& Govindjee, Energy conversion in photosynthetic bacteria.
\emph{Photosynth. Res}., \textbf{127}(2), 257--271, (2016).

\bibitem{Maroti_2019a}
Maróti, P. Chemical rescue of H\textsuperscript{+} delivery in proton
transfer mutants of reaction center of photosynthetic bacteria.
\emph{Biochim. Biophys Acta} -- Bioenergetics, \textbf{1860}, 317--324,
(2019a).

\bibitem{Maroti_2019b}
Maróti, P. Thermodynamic View of Proton Activated Electron Transfer in
the Reaction Center of Photosynthetic Bacteria. \emph{J. Phys. Chem}.
\emph{B,} \textbf{123}, 5463-5473, (2019b).

\bibitem{Franck_and_Teller_1938}
Franck, J. \& Teller, E. Migration and Photochemical Action of
Excitation Energy in Crystals. \emph{J. Chem. Phys}., \textbf{6},
861-872, (1938).

\bibitem{Niederman_2016}
Niederman, R.A. Development and dynamics of the photosynthetic apparatus
in purple phototrophic bacteria. \emph{Biochim. Biophys. Acta,}
\textbf{1857}, 232-246, (2016).

\bibitem{Vredenberg_and_Duysens_1963}
Vredenberg, W.J. \& Duysens, L.N.M. Transfer and trapping of excitation
energy from bacteriochlorophyll to a reaction center during bacterial
photosynthesis. \emph{Nature}, \textbf{197}, 355--357, (1963).

\bibitem{Joliot_et_al_1973}
Joliot, P., Bennoun, P. \& Joliot, A. New evidence supporting energy
transfer between photosynthetic units. \emph{Biochim Biophys Acta,}
\textbf{305}, 317-328, (1973).

\bibitem{Paillotin_1976}
Paillotin, G. Capture frequency of excitations and energy transfer
between photosynthetic units in the photosystem II. \emph{J. Theor.
Biol}., \textbf{58}, 219-235, (1976).

\bibitem{Lavergne_and_Trissl_1995}
Lavergne, J. \& Trissl, H.W. Theory of fluorescence induction in
photosystem-II---derivation of analytical expressions in a model
including exciton-radical-pair equilibrium and restricted
energy-transfer between photosynthetic units. \emph{Biophys. J}.,
\textbf{68}, 2474--2492, (1995).

\bibitem{de_Rivoyre_et_al_2010}
de Rivoyre, M., Ginet, N., Bouyer, P. \& Lavergne, J. Excitation
transfer connectivity in different purple bacteria: a theoretical and
experimental study. \emph{Biochim. Biophys. Acta,} \textbf{1797},
1780-1794, (2010).

\bibitem{Trissl_1996}
Trissl, H.W. Antenna organization in purple bacteria investigated by
means of fluorescence induction curves. \emph{Photosynth. Res}.,
\textbf{47,} 175--185, (1996).

\bibitem{Den_Hollander_et_al_1983}
Den Hollander, W.T.F., Bakker, J.G.C. \& van Grondelle, R. Trapping,
loss and annihilation of excitations in a photosynthetic system. I.
Theoretical aspects. \emph{Biochim. Biophys. Acta,} \textbf{725},
492-507, (1983).

\bibitem{Fassioli_et_al_2009}
Fassioli, F., Olaya-Castro, A., Scheuring, S., Sturgis, J.N. \& Johnson,
N.F. Energy transfer in light-adapted photosynthetic membranes: from
active to saturated photosynthesis. \emph{Biophys. J}., \textbf{97,}
2464--2473, (2009).

\bibitem{Sebban_and_Barbet_1985}
Sebban, P. \& Barbet, J.C. Simulation of the energy migration in the
antenna of purple bacteria by using the Monte Carlo method.
\emph{Photobiochem. Photobiophys}., \textbf{9}, 167-175, (1985).

\bibitem{Asztalos_et_al_2015}
Asztalos, E., Sipka, G. \& Maróti, P. Fluorescence relaxation in intact
cells of photosynthetic bacteria: donor and acceptor side limitations of
reopening of the reaction center. \emph{Photosynth. Res.,}
\textbf{124}(1), 31-44, (2015).

\bibitem{Maroti_2016}
Maróti, P. Induction and relaxation of bacteriochlorophyll fluorescence
in photosynthetic bacteria. In: Pessarakli M. (ed): \emph{Handbook of
Photosynthesis}, 3rd ed. Pp. 463-483. CRC Press, Boca Raton -- London --
New York (2016).

\bibitem{Kupper_et_al_2019}
Küpper, H., Benedikty, Z., Morina, F., Andresen, E., Mishra, A. \&
Trtílek, M. Analysis of OJIP Chlorophyll Fluorescence Kinetics and
Q\textsubscript{A} Reoxidation Kinetics by Direct Fast Imaging.
\emph{Plant. Physiol}., \textbf{179}, 369--381, (2019).

\bibitem{Niederman_2013}
Niederman, R.A. Membrane development in purple photosynthetic bacteria
in response to alterations in light intensity and oxygen tension.
\emph{Photosynth Res}, \textbf{116}, 333--348, (2013).

\bibitem{site_perc} Feng. X., Deng Y. and H.W.J. Bl\"ote H.W.J., Percolation transitions in two dimensions,
\emph{Phys. Rev. E} \textbf{78}, 031136 (2008).

\bibitem{site_perc1} Jacobsen J. L., High-precision percolation thresholds and Potts-model critical manifolds from graph polynomials, \emph{J. Phys. A: Math. Theor.} \textbf{47}, 135001 (2014).

\bibitem{Wraight_and_Clayton_1974}
Wraight, C.A. \& Clayton, R.K. The absolute quantum efficiency of
bacteriochlorophyll photooxidation in reaction centers of
\emph{Rhodopseudomonas sphaeroides}. \emph{Biochim. Biophys. Acta,}
\textbf{333}, 246-260, (1974).

\bibitem{Timpmann_et_al_1993}
Timpmann, K., Zhang, F.G., Freiberg, A. \& Sundstrom, V. Detrapping of
excitation energy from the reaction centre in the photosynthetic purple
bacterium \emph{Rhodospirillum rubrum}. \emph{Biochim. Biophys. Acta}.,
\textbf{1183}, 185-193, (1993).


\bibitem{Scheuring_et_al_2005}
Scheuring, S., Levy, D. \& Rigaud, J.L. Watching the components of
photosynthetic bacterial membranes and their in situ organisation by
atomic force microscopy. \emph{Biochim. Biophys. Acta,} \textbf{1712},
109--127, (2005).

\bibitem{Liu_and_Scheuring_2013}
\href{https://www.cell.com/trends/plant-science/fulltext/S1360-1385(13)00038-1?_returnURL=https\%3A\%2F\%2Flinkinghub.elsevier.com\%2Fretrieve\%2Fpii\%2FS1360138513000381\%3Fshowall\%3Dtrue}{Liu},
L. N. \&
\href{https://www.cell.com/trends/plant-science/fulltext/S1360-1385(13)00038-1?_returnURL=https\%3A\%2F\%2Flinkinghub.elsevier.com\%2Fretrieve\%2Fpii\%2FS1360138513000381\%3Fshowall\%3Dtrue}{Scheuring},
S. Investigation of photosynthetic membrane structure using atomic force
microscopy. \emph{Trend in Plant Science}, \textbf{18}(5), 277-286,
(2013).

\bibitem{Xin_et_al_2018}
Xin, Y., Shi, Y., Niu, T., Wang, Q., Niu, W., Huang, X., Ding, W., Yang,
L., Blankenship, R. E., Xu, X. \& Sun, F. Cryo-EM structure of the RC-LH
core complex from an early branching photosynthetic prokaryote.
\emph{Nature Communications} \textbf{9}:1568, (2018).

\bibitem{Kramer_and_Rodriguez_2017}
Kramer, T. \& Rodriguez, M. Two-dimensional electronic spectra of the
photosynthetic apparatus of green sulfur bacteria. \emph{Scientific
Reports,} \textbf{7}:45245, (2017).

\bibitem{Panitchayangkoon_et_al_2010}
Panitchayangkoon, G., Hayes, D., Fransted, K.A., Caram, J.R., Harel, E.,
Wen, J., Blankenship, R.E. \& Engel, G.S. Long-lived quantum coherence
in photosynthetic complexes at physiological temperature. ~\emph{PNAS},
\textbf{107}, 12766-12770, (2010).

\bibitem{Ishizaki_and_Fleming_2012}
Ishizaki, A. \& Fleming, G. R. Quantum Coherence in Photosynthetic Light
Harvesting.
\href{https://www.annualreviews.org/journal/conmatphys}{\emph{Annual
Review of Condensed Matter Physics},~}\textbf{3}, 333-361, (2012).

\bibitem{Ball_2018}
Ball, P. ``Is photosynthesis quantum-ish?'' \emph{Physics World,}
\textbf{31} (4), 44, (2018).

\bibitem{Comayras_et_al_2005}
Comayras, F., Jungas, C. \& Lavergne, J. Functional consequences of the
organization of the photosynthetic apparatus in \emph{Rhodobacter
sphaeroides}. I. Quinone domains and excitation transfer in
chromatophores and reaction center-antenna complexes. \emph{J. Biol.
Chem}., \textbf{280}, 11203--11213, (2005).

\bibitem{Olsen_et_al_2017}
\href{https://www.ncbi.nlm.nih.gov/pubmed/?term=Olsen\%20JD\%5BAuthor\%5D\&cauthor=true\&cauthor_uid=28130884}{Olsen},~J.D.,~\href{https://www.ncbi.nlm.nih.gov/pubmed/?term=Martin\%20EC\%5BAuthor\%5D\&cauthor=true\&cauthor_uid=28130884}{Martin},
E.C. \&
~\href{https://www.ncbi.nlm.nih.gov/pubmed/?term=Hunter\%20CN\%5BAuthor\%5D\&cauthor=true\&cauthor_uid=28130884}{Hunter},
C.N. The PufX quinone channel enables the light-harvesting 1 antenna to
bind more carotenoids for light collection and photoprotection.
\href{https://www.ncbi.nlm.nih.gov/pmc/articles/PMC5347945/}{\emph{FEBS
Lett}}., \textbf{591}(4), 573--580, (2017).

\bibitem{Sipka_and_Maroti_2018}
Sipka, G. \& Maróti P. Photoprotection in intact cells of photosynthetic
bacteria: quenching of bacteriochlorophyll fluorescence by carotenoid
triplets. \emph{Photosynth. Res.} \textbf{136,} 17--30, (2018).

\bibitem{Sipka_et_al_2018}
Sipka, G., Kis, M., Smart, J.L. \& Maróti, P. Fluorescence induction of
photosynthetic bacteria. \emph{Photosynthetica,} \textbf{56} (1),
125-131, (2018).

\bibitem{Chi_et_al_2015}
Chi, S.C., Mothersole, D.J., Dilbeck P. et al.: Assembly of functional
photosystem complexes in \emph{Rhodobacter sphaeroides} incorporating
carotenoids from the spirilloxanthin pathway. \emph{Biochim. Biophys.
Acta,} \textbf{1847}, 189-201, (2015).

\bibitem{Sambrook_et_al_1989}
Sambrook, J., Fritsch, E.F. \& Maniatis, T. Molecular cloning: a
laboratory manual, 2nd ed. Page A.1. Cold Spring Harbor Laboratory
Press, New York (1989).

\bibitem{Sistrom_1977}
Siström, W.R. Transfer of chromosomal genes mediated by plasmid r68.45
in \emph{Rhodopseudomonas sphaeroides}. \emph{J. Bacteriol}.,
\textbf{131}, 526-532, (1977).

\bibitem{Donohue_and_Kaplan_1991}
Donohue, T.J. \& Kaplan, S. Genetic techniques in rhodospirillaceae.
\emph{Methods Enzymol}., \textbf{204}, 459-485, (1991).

\bibitem{Kis_et_al_2014}
Kis, M., Asztalos, E., Sipka, G. \& Maróti, P. Assembly of
photosynthetic apparatus in \emph{Rhodobacter sphaeroides} as revealed
by functional assessments at different growth phases and in synchronized
and greening cells. \emph{Photosynth. Res}., \textbf{122}, 261-273,
(2014).

\bibitem{Bina_et_al_2010}
Bína, D., Litvín R. \& Vácha, F. Absorbance changes accompanying the
fast fluorescence induction in the purple bacterium \emph{Rhodobacter
sphaeroides}. \emph{Photosynth. Res}., \textbf{105}, 115-121, (2010).

\bibitem{Kocsis_et_al_2010}
Kocsis, P., Asztalos, E., Gingl, Z. \& Maróti, P. Kinetic
bacteriochlorophyll fluorometer. \emph{Photosynth. Res}., \textbf{105},
73-82, (2010).

\end{thebibliography}
\end{document}